\documentclass[onecolumn,showpacs,aps,epsfig,nofootinbib]{revtex4}
\usepackage{graphicx}
\usepackage{epstopdf}
\usepackage{latexsym}
\usepackage{graphicx,times}
\usepackage{natbib}
\usepackage{amssymb,amsmath}
\usepackage{color}
\usepackage{colordvi}
\usepackage[center]{subfigure}

\begin{document}

%%%%%%%%%%%%%%%%%%%%%%%%%%%%%%%%%%%%%%%%%%%%%%%%%%%%%%%%%%%%%%%
 \newcommand{\bq}{\begin{equation}}
 \newcommand{\eq}{\end{equation}}
 \newcommand{\bqn}{\begin{eqnarray}}
 \newcommand{\eqn}{\end{eqnarray}}
 \newcommand{\nb}{\nonumber}
 \newcommand{\lb}{\label}
 \newcommand{\tc}{\textcolor{black}}
\newcommand{\PRL}{Phys. Rev. Lett.}
\newcommand{\PL}{Phys. Lett.}
\newcommand{\PR}{Phys. Rev.}
\newcommand{\CQG}{Class. Quantum Grav.}
 %%%%%%%%%%%%%%%%%%%%%%%%%%%%%%%%%%%%%%%%%%%%%%%%%%%%%%%%%%%%%%%

%\title{Probing the statistical properties of CMB $B$-mode polarization through Minkowski Functionals}

%\author{ Larissa Santos, Wen Zhao, Kai Wang}
% \affiliation{CAS Key Laboratory for Researches in Galaxies and Cosmology, Department of Astronomy, University of Science and Technology of China, Chinese Academy of Sciences, Hefei, Anhui 230026}
%

\title{Statistical imprints of CMB $B$-type polarization leakage in an incomplete sky survey analysis}

\author{Larissa Santos}
\email[]{larissa@ustc.edu.cn}
\author{Kai Wang}
\author{Yangrui Hu}
\author{Wenjuan Fang}
\author{Wen Zhao}
\email[]{wzhao7@ustc.edu.cn} \affiliation{CAS Key Laboratory for
Researches in Galaxies and Cosmology, Department of Astronomy,
University of Science and Technology of China, Chinese Academy of
Sciences, Hefei, Anhui 230026, China}

\date{\today}

\begin{abstract}

One of the main goals of modern cosmology is to search for
primordial gravitational waves by looking on their imprints in the
$B$-type polarization in the cosmic microwave background
radiation. However, this signal is contaminated by various
sources, including cosmic weak lensing, foreground radiations,
instrumental noises, as well as the $E$-to-$B$ leakage caused by
the partial sky surveys, which should be well understood to avoid
the misinterpretation of the observed data. In this paper, we
adopt the $E$/$B$ decomposition method suggested by Smith in 2006,
and study the imprints of $E$-to-$B$ leakage residuals in the
constructed $B$-type polarization maps, $\mathcal{B}(\hat{n})$, by
employing various statistical tools. We find that the effects of
$E$-to-$B$ leakage are negligible for the $\mathcal{B}$-mode power
spectrum, as well as the skewness and kurtosis analyses of
$\mathcal{B}$-maps. However, if employing the morphological
statistical tools, including Minkowski functionals and/or Betti
numbers, we find the effect of leakage can be detected at very
high confidence level, which shows that in the morphological
analysis, the leakage can play a significant role as a contaminant
for measuring the primordial $B$-mode signal and must be taken
into account for a correct explanation of the data.

\end{abstract}


\maketitle

\section{Introduction}

The temperature and polarization anisotropies of the cosmic
microwave background (CMB) radiation are powerful cosmological
observables to understand the physics of the early universe.
During the past decades, much effort has been made to characterize
the CMB, including three satellite generations: the cosmic
background explorer (COBE), the Wilkinson microwave anisotropy
probe (WMAP) and Planck. These experiments were mainly devoted to
measure the CMB temperature anisotropies, precisely mapping these
tiny fluctuations in the sky \cite{wmap9,planck2015}. However,
according to the standard cosmological model, the full information
of CMB is encoded in the statistical properties of both
temperature,  $T(\hat{\gamma})$, and the linear polarizations,
described by the stocks parameters, $Q(\hat{\gamma})$ and
$U(\hat{\gamma})$, which are defined with respect to a fixed
coordinate system in the sky, and depend on the choice of
coordinate system. It is then convenient to decompose the linear
polarization into the curl-free ($E$-mode) and divergence-free
($B$-mode) components since they have the advantage of being
rotationally invariant
\cite{zaldarriaga-b-mode,kamionkowski-b-mode}.

Recent results from the Planck satellite also show precise
measurements of the $E$-mode field \cite{planck-e-mode}. In the
standard cosmological model, both CMB temperature and $E$-mode
anisotropies are mainly generated by the primordial density
perturbations. In addition, the auto-correlation power spectra
$C_{\ell}^{TT}$ and $C_{\ell}^{EE}$ and the cross-correlation
power spectrum $C_{\ell}^{TE}$ sensitively depend on the
cosmological parameters and cosmological models. Thus, the
observations of WMAP and Planck satellites on these spectra have
tightly constrained most cosmological parameters and inflationary
parameters. However, the $B$-mode polarization encodes quite
different cosmological information: In the large scales, the
$B$-mode polarization is supported to be generated by the
primordial gravitational waves
\cite{zaldarriaga-b-mode,kamionkowski-b-mode,gw-b-mode}, which is
the smoking-gun evidence of inflation \cite{pgw}. In the small
scales, the $B$-mode is mainly produced by the deflection of the
CMB photons by cluster of galaxies during their travel between the
last scattering surface and the observer, known as the CMB lensing
\cite{zaldarriaga-lensing,hu-lensing,lewis-review}. During the
past few years, many ground-based experiments were designed to
measure the $B$-mode polarization signal, such as SPTPol
\cite{sptpol}, POLARBEAR \cite{polarbear}, ACTPol \cite{actpol},
BICEP2 and Keck Array \cite{bicep2-1,bicep2-2,bicep2-3,bicep2-4}.
These experiments, including the Planck satellite
\cite{planck-b-mode}, have detected the lensed $B$-mode signal in
the high multipole range at the high confidence level. Thus, the
detection of the primordial $B$-mode signal is then the main goal
of future CMB experiments \cite{cmb-taskforce}. Among them we can
cite BICEP3, AdvACT, CLASS, Simons Array, SPT-3G, C-BASS, QUIJOTE,
EBEX, QUBIC, QUIET, PIPER, Spider, LSPE, et al.
\cite{ground-based} as ground-based experiments, and  LiteBIRD
\cite{litebird}, CMBPOL \cite{cmbpol}, COrE \cite{core}, PRISM
\cite{prism}, PIXIE \cite{pixie} as the next satellite generation
of CMB experiments.

In the real observations, the detection of cosmological $B$-mode
is limited by various contaminations, including instrumental
noises, instrumental systematical errors \cite{task}, as well as
the polarized foregrounds \cite{foregrounds,planck-foregrounds}.
Another source of contamination is the so-called $E$-$B$ mixture,
which arises from an incomplete sky analysis of the CMB
polarization signals \cite{ebmixture1}. Our ability to decompose
CMB polarization signal in a partial sky coverage is crucial,
since even for satellite missions the presence of non-cosmological
contaminations must be masked out. So, this mixture exists in all
the CMB polarization analysis, and could become the dominant
contamination for the detection of primordial gravitational waves
\cite{ebmixture1}. In order to solve this problem, numerous
practical methods have been developed to separate the $E$-mode and
$B$-mode in the partial sky
\cite{ebmixture0,ebmixture2,ebmixture3,ebmixture4,ebmixture5}.
However, most of these methods suffer from one or several of the
following drawbacks - they are slow in practice, they are
difficult to realize in pixel space, and/or they lead to partial
information loss. For instance, the method suggested by Bunn et
al. (this method is adopted by BICEP2/Keck Array collaboration
\cite{bicep2-1,bicep2-4}) involves constructing an eigenbasis, and
it has a high computational cost \cite{ebmixture0,leistedt}. Among
all these methods, the methods proposed in
\cite{ebmixture3,ebmixture4,ebmixture5} are based on a common
algebraic framework of the so-called $\chi$-fields. Therefore,
they are fast and can be efficiently applied to high resolution
maps due to the use of a fast spherical harmonics transformation.
However, it is important to mention that the residual of the
$E$-to-$B$ leakage is unavoidable even if the proper method is
applied in the data analysis \cite{Wang}. In the previous work
\cite{ferte2013}, the authors carefully compared these three
methods, and they found that although all of them allow a
significant reduction of the level of the $E$-to-$B$ leakage, the
method of Smith \cite{ebmixture3} ensures the smallest error bars
in all experimental configurations and leads to the smallest
leakage residuals. For these reasons, in the present article, we
shall focus on this method due to our limited computational
resources, and study the influence of residuals of $E$-to-$B$
leakage on the CMB $B$-mode polarization.

In most previous works \cite{ebmixture3,ferte2013}, the authors
focused only on the power spectrum of the $B$-type polarization
and that the $E$-to-$B$ leakage is tiny in the Smith's $E$/$B$
decomposition method. However, the $B$-mode is a highly
non-gaussian field due to the different kind of components in the
observed maps. Thus, in addition to the power spectrum, various
non-gaussian statistical tools are also helpful to separate the
different components and/or constrain the cosmological parameters.
In the previous work \cite{larissa}, we have applied the Minkowski
functionals (MFs) to quantify the deviation from Gaussianity of
$B$-mode maps, and studied the effects of instrumental noises, the
CMB mask, and the influence of foreground residuals. As an
extension of this work, in the present paper, by applying a
similar analysis we will focus on characterizing the imprint of
the $E$-$B$ mixture residual in the lensed $B$-mode map. As well
known, MFs characterize the morphological properties of convex,
compact sets in an $n$-dimensional space
\cite{minkowski1903,gott1990,schneider1993,mecke1994,schmalzing1997,schmalzing1998,wizitzki1998}.
Moreover, the $E$-to-$B$ leakage has completely different
morphological properties from the stochastic $B$-mode generated by
cosmological sources. We expect the influence of the $E$-to-$B$
leakage to be significant in the MFs analysis, even if the
amplitude of this leakage is small. As the a complementary
analysis, in this paper we shall also apply other statistical
tools, including the Betti numbers, skewness and kurtosis, to
investigate the imprints of $E$-to-$B$ leakage. Throughout this
paper, in order to focus on the effect of $E$-to-$B$ leakage
residuals, we shall consider the case with only cosmic variance
and ignore the effects of the other $B$-mode sources, including
the instrumental noises and foregrounds, which have been detail
studied in the previous work \cite{larissa}.

The paper is divided as follows: In Sec. \ref{$E$- and $B$-mode
polarization of CMB} we briefly introduce the theory of CMB
polarization. In Sec. \ref{$E$- and $B$-mode decomposition in
partial sky}, we derive the equations needed to obtain the $E$/$B$
decomposition in a partial sky analysis. Sec. \ref{Minkowski
Functionals as the statistic} is divided in 3 subsections: Sec.
\ref{Minkowski Functionals} is devoted to the introduction of the
MF statistics, Sec. \ref{method} details the methodology used in
this paper, and Sec. \ref{Results} describes the results obtained
for the MFs. In Sec. \ref{Other statistics}, we use other
statistics in the analysis for completeness, being the Betti
numbers described in Sec. \ref{Betti numbers}, and both skewness
and kurtosis explained in Sec. \ref{Skewness and kurtosis}.
Finally, in Sec. \ref{Conclusions}, we draw our conclusions.

\section{$E$- and $B$-mode polarization of CMB}
\label{$E$- and $B$-mode polarization of CMB}

The linearly polarized CMB polarization field is completely
described by two Stokes parameters, $Q$ and $U$ that can be
combined into a spin-(2) and spin-(-2) fields
$P_{\pm}(\hat{n})=Q(\hat{n})\pm iU(\hat{n})$. For full sky, the
spin fields can be expanded over spin-weighted harmonic functions
basis as follows \cite{seljak1996}:
\begin{equation}
P_{\pm}(\hat{n})=\sum_{\ell m} a_{\pm2,\ell m}~_{\pm 2}Y_{\ell m}(\hat{n}).
\end{equation}
Alternatively, the polarization field can be written as the
curl-free $E$, and divergence-free $B$ components, which are
defined in terms of the coefficients $a_{\pm 2,\ell m}$ in the
harmonic space as
\begin{eqnarray}
\label{eq_elm}
E_{\ell m}\equiv -\frac{1}{2}[a_{2,\ell m}+a_{-2,\ell m}], ~~~~~
B_{\ell m}\equiv -\frac{1}{2i}[a_{2,\ell m}-a_{-2,\ell m}].
\end{eqnarray}

In the same way as the temperature field, we can define the $E(\hat{n})$ and $B(\hat{n})$ polarization sky maps in terms of spherical harmonics,

 \begin{eqnarray}
 E(\hat{n})\equiv \sum_{\ell m}E_{\ell m} Y_{\ell m}(\hat{n}),~~~~~
 B(\hat{n})\equiv \sum_{\ell m}B_{\ell m} Y_{\ell m}(\hat{n}).
 \end{eqnarray}
The power spectra can then be written as
 \begin{eqnarray}
 C_{\ell}^{EE}&\equiv&\frac{1}{2\ell+1}\sum_m\langle E_{\ell m} E_{\ell m}^* \rangle, \\
 C_{\ell}^{BB}&\equiv&\frac{1}{2\ell+1}\sum_m\langle B_{\ell m} B_{\ell m}^* \rangle,
 \end{eqnarray}
where the brackets denote the average over all realizations. For a
Gaussian field, all the statistical properties can be obtained by
analyzing the second-order power spectra. It is important to
emphasize, however, that here we are dealing with a highly
non-Gaussian lensed $B$-map due to the contribution of CMB lensing
to the final map \cite{larissa}. In this case, different
statistics, as, for example, the MFs, are necessary to describe
the field. Moreover, we notice that the $E$/$B$ decomposition in a
full sky analysis is straightforward, providing a direct link to
the primordial cosmological perturbations, especially the GWs
imprint on the CMB $B$-mode polarization.  Nevertheless, we know
that Galactic foregrounds are present even in full sky surveys,
and they should be masked to reduce the contaminations. So, in the
realistic case, we must derive the $E$-type and $B$-type maps from
the incomplete $Q$ and $U$ observables.

\section{$E$- and $B$-mode decomposition in partial sky}
\label{$E$- and $B$-mode decomposition in partial sky}

If the polarization fields are not measured in full sky, but on a
fraction only, Eqs.(\ref{eq_elm}) cannot be derived directly. So,
the separation of pure $E$- and $B$-mode from the observed $Q$ and
$U$ maps is not trivial, due to the existence of ambiguous mode
\cite{ebmixture0}, which can be successfully avoided by different
ways
\cite{ebmixture0,ebmixture2,ebmixture3,ebmixture4,ebmixture5}. As
mentioned above, in this paper, we adopt the method suggested in
\cite{ebmixture3}, which is based on the algebraic framework of
the so-called $\chi$-field (denoted as $\mathcal{B}$-field in the
present paper).

For the full-sky observations, according to
\cite{zaldarriaga-seljak}, we can define a new set of fields
$\mathcal{E}$ and $\mathcal{B}$ from the polarization fields $Q$
and $U$ as follows:
\begin{eqnarray}
\label{pseudo_E}
\mathcal{E}(\hat{n}) = -\frac{1}{2}[\bar{\eth}\bar{\eth}P_{+}(\hat{n}) + \eth \eth P_{-}(\hat{n})], \\
\label{pseudo_B}
\mathcal{B}(\hat{n}) = -\frac{1}{2i}[\bar{\eth}\bar{\eth}P_{+}(\hat{n}) - \eth \eth P_{-}(\hat{n})],
\end{eqnarray}
where $\eth (\bar{\eth})$ corresponds to the spin-raising
(lowering) operator for an arbitrary function $f$ with spin $s$,
\begin{eqnarray}
\eth f \equiv -\sin^s \theta \left(\frac{\partial}{\partial \theta} + \frac{i}{\sin \theta}\frac{\partial}{\partial \phi}\right) (f \sin^{-s}\theta), \\
\bar{\eth} f \equiv -\sin^{-s} \theta \left(\frac{\partial}{\partial \theta} - \frac{i}{\sin \theta}\frac{\partial}{\partial \phi}\right) (f \sin^s\theta).
\end{eqnarray}
From the definition, we know that the new map $\mathcal{E}$ in Eq.
(\ref{pseudo_E}) is the standard scalar field, and $\mathcal{B}$
in (\ref{pseudo_B}) is the pseudo-scalar field in the
two-dimensional sphere. Thus, they can be expanded in the
spherical harmonics as follows,
\begin{eqnarray}
\label{pseudo_E_harm}
\mathcal{E}(\hat{n}) \equiv \sum_{\ell m}\mathcal{E}_{\ell m}Y_{\ell m}(\hat{n}),~~~~~~
\label{pseudo_B_harm}
\mathcal{B}(\hat{n}) \equiv \sum_{\ell m}\mathcal{B}_{\ell m}Y_{\ell m}(\hat{n}),
\end{eqnarray}
where the expanding coefficients are
\begin{eqnarray}
 \label{Elm_2}
\mathcal{E}_{\ell m} =  \int  \mathcal{E}(\hat{n})Y_{\ell m}^*(\hat{n})d\hat{n},~~~~~
\mathcal{B}_{\ell m} =  \int  \mathcal{B}(\hat{n})Y_{\ell m}^*(\hat{n})d\hat{n}.
\end{eqnarray}
These coefficients are related to the regular multipoles $E_{\ell m}$ and $B_{\ell m}$ by
\begin{equation}
 \label{Elm_pseudo}
\mathcal{E}_{\ell m}=N_{\ell,2}E_{\ell m},~~~~~\mathcal{B}_{\ell m}=N_{\ell,2}B_{\ell m},
\end{equation}
where we have $N_{\ell ,s}=\sqrt {{(\ell +s)!}/{(\ell -s)!}}$ \cite{zaldarriaga-seljak}. The corresponding power spectra are then obtained as
 \begin{eqnarray}
 \label{pseudo_cl}
 C_{\ell}^{\mathcal{EE}}&\equiv&\frac{1}{2\ell+1}\sum_m\langle \mathcal{E}_{\ell m} \mathcal{E}_{\ell m}^* \rangle=N_{\ell ,2}^2 C_{\ell}^{EE}, \\
 C_{\ell}^{\mathcal{BB}}&\equiv&\frac{1}{2\ell+1}\sum_m\langle \mathcal{B}_{\ell m} \mathcal{B}_{\ell m}^* \rangle=N_{\ell ,2}^2 C_{\ell}^{BB}.
 \end{eqnarray}

%With this alternative method to get the $B$-type polarization, now we can deal with the case of partial sky experiments.

%With the help of the mask function $W(\hat{n})$, the pseudo multipoles become \cite{efstthiou2004}

Considering the mask window function $W(\hat{n})$, the masked
$B$-type polarization map becomes $\mathcal{B}W(\hat{n})$, and the
pseudo multipole coefficients can be defined as follows
\cite{efstthiou2004},
\begin{eqnarray}
%\label{pseudo_Elm}
%\tilde{\mathcal E}_{\ell m} = \int d\hat{n} W(\hat{n}) \mathcal{E}(\hat{n})Y_{\ell m}^*(\hat{n}), \\
\label{pseudo_Blm}
\tilde{\mathcal B}_{\ell m} = \int d\hat{n}W(\hat{n}) \mathcal{B}(\hat{n})Y_{\ell m}^*(\hat{n}).
\end{eqnarray}
So, the pure $B$-type map $\mathcal{B}W(\hat{n})$ and the coefficient set $\tilde{\mathcal B}_{\ell m}$ are mathematically equivalent by definition. In the previous work \cite{ebmixture4}, we have developed to method to directly construct the pure map $\mathcal{B}W(\hat{n})$ from the masked observables $Q$ and $U$. While in the method \cite{ebmixture3}, we should first construct the pure coefficients $\tilde{\mathcal B}_{\ell m}$ and then translate them into the map $\mathcal{B}W(\hat{n})$. However, it is important to emphasize that in both methods, it is impossible to construct the pure $B$-maps directly. In this paper, we shall adopt the latter method, which is briefly reviewed as follows.

%By comparing three different methods developed to extract the $E$ and $B$ signals from $Q$ and $U$, \cite{ferte2013} concluded that the Smith and Zaldarriaga method (SZ, hereafter) \cite{ebmixture3} is the most efficient since it significantly reduces the $E$-to-$B$ leakage, and at the same time ensures the smallest error bars. Recently, \cite{leistedt} demonstrated a reduction of the leakage using their wavelet estimator over the standard harmonic space approaches. However, due to our limited computational resources, we followed the previous result by  \cite{ferte2013} and used the SZ method throughout this paper, which starts by the definition of pure pseudo multipoles

In this method, the concept of pure pseudo-multipoles is put forward and defined as,
\begin{align}
\label{pureb} \mathcal{B}_{\ell m}^{\rm
pure}\equiv-\frac{1}{2i}\int d\hat{n}\left\{P_{+}(\hat
n)\left[\bar {\eth} \bar{\eth}\left(W(\hat{n})Y_{\ell
m}(\hat{n})\right)\right]^\ast-P_{-}(\hat n)\left[ {\eth}
{\eth}\left(W(\hat{n})Y_{\ell m}(\hat{n})\right)\right]^\ast
\right\}.
\end{align}
It can be proved that this definition is equivalent to the Eq.
(\ref{pseudo_Blm}) \cite{ebmixture3}, which shows that in
principle the pure pseudo-multipole method can successfully
extract the pure $B$-type polarization signal and avoid the
$E$-$B$ mixing part. To calculate the expression of Eq.
(\ref{pureb}), we use the property of spin raising and lowering
operators and obtain that \cite{ferte2013}
\begin{align}\label{purebb}
\mathcal{B}_{\ell m}^{\rm pure}=-\frac{1}{2i}\int d \hat{n} \bigg{[}
& P_{+}\bigg{(} \left(\bar {\eth} \bar{\eth}W\right) Y_{\ell m}^{\ast}+2N_{\ell,1} \left(\bar{\eth } W\right)  \left(\sideset{_{1}}{_{\ell m}^{\ast}}{\mathop{\mathrm{Y}}}\right)+N_{\ell,2} W \left(\sideset{_{2}}{_{\ell m}^{\ast}}{\mathop{\mathrm{Y}}}\right)\bigg{)} \nonumber \\
 -& P_{-}\bigg ( \left({\eth}{\eth}W \right)Y_{\ell m}^{\ast}-2N_{\ell,1}\left({\eth } W\right ) \left( \sideset{_{-1}}{_{\ell m}^{\ast}}{\mathop{\mathrm{Y}}}\right)+N_{\ell,2} W\left ( \sideset{_{-2}}{_{\ell m}^{\ast}}{\mathop{\mathrm{Y}}}\right)\bigg )
\bigg{]}.
\end{align}
This expression is used in the following calculations.

Note that, in this method, the first and second derivatives of the
window function $W$ are used. If we adopt the top-hat window
function, these derivatives diverge in the numerical computations.
So, in practice, we should properly smooth the window function to
avoid the divergence at the observed patch boundaries. Thus, an
appropriate sky apodization will play an important role in
suppressing the $E$-to-$B$ leakage. In the previous work
\cite{Wang}, we carefully compare the leakage residual and the
information loss for different smoothing methods, and found that
the Gaussian smoothing method presented by \cite{kim2011} induces
the smallest leakage in the final $B$-map \footnote{In the
previous works \cite{ebmixture3}, the authors discovered the
optimal smoothing function to minimize the errors of the
constructed power spectra $C_{\ell}^{BB}$, which is dependent of
the multipole $\ell$. However, different from them, in this paper
we shall only focus on the statistical properties of the
constructed $\mathcal{B}$-map. So, we choose the window function
for different aim, which should minimize the $E$-$B$ leakage and
reduce the information loss.}. The Gaussian smoothing kernel used
to smooth the edges of $W$ is defined as \cite{kim2011},
\begin{equation}\label{WF}
W_i=\left \{
\begin{aligned}
&\int_{-\infty }^{\delta_i-\frac{\delta_c}{2}} \frac{1}{\sqrt{2\pi \sigma^2}}\exp\bigg(-\frac{x^2}{2\sigma^2}\bigg)dx=\frac{1}{2}+\frac{1}{2}{\rm erf}\bigg(\frac{\delta_i-\frac{\delta_c}{2}}{\sqrt{2}\sigma }\bigg)  &\delta_{i}<\delta_{c}\\
&1  &\delta_{i}>\delta_{c}\\
\end{aligned}
\right.
\end{equation}
where $ \sigma={\theta_{\rm F}}/{\sqrt{8\ln2}} $ and $\theta_{\rm
F}$ denotes the full width at half maximum of the smoothing
kernel. $\delta_i$ the smallest angular distance between the
$i$-th observed pixel and the boundary of the mask. $\delta_c$ is
an adjustable parameter referred as the apodization length. Let $
\beta $ denotes the jump range at $\delta_i=\delta_c $ and
$\delta_i=0 $, which is (for details see \cite{kim2011}):
\begin{equation}
\label{betha}
\beta = \frac{1}{2} - \frac{1}{2} \text{erf} \left(\frac{\frac{\delta_c}{2}}{\sqrt 2 \sigma}\right).
\end{equation}
$\beta $ is a small and adjustable parameter, which must be chosen
by investigating the reconstruction numerical accuracy of a
smoothed mask performed using the HEALPix package \cite{Wang}.
Here we should mention that performing a spherical harmonic
transformation of the foreground mask and its later reconstruction
by inverse transformation leads to an oscillation pattern around
jump discontinuities (i.e. the  reconstructed mask has non-zero
values where they were originally zero), called the Gibbs
phenomenon. The discrepancy between the original mask and the
reconstructed one must be corrected by choosing a window function
in which multipoles higher than the truncation point are
suppressed.

%\begin{equation}
%\label{WF}
%W =\left \{ \begin{array} {ll}
%\frac{1}{2} + \frac{1}{2} \text{erf} \left(\frac{\delta_i - \frac{\delta_c}{2}}{\sqrt 2 \sigma}\right) \quad \delta_i < \delta_c,\\
%1 \quad  \quad \quad \quad \quad \quad ~\quad  \quad \delta_i >\delta_c,
%\end{array} \right.
%\end{equation}
%where $\delta_i$ the smallest angular distance between the $i$-th observed pixel and the boundary of the mask. $\delta_c$ is an adjustable parameter referred as the apodization length. Performing a spherical harmonic transformation of the foreground mask and its later reconstruction by inverse transformation leads to an oscillation pattern around jump discontinuities (i.e. the  reconstructed mask has non-zero values where they were originally zero), called the Gibbs phenomenon. The discrepancy between the original mask and the reconstructed one must be corrected by choosing a window function in which multipoles higher than the truncation point are suppressed. A new parameter, $\beta$, is then defined as (for details see \cite{kim2011}):

%$\beta$ has a very small value that must be chosen by investigating the reconstruction numerical accuracy of a smoothed mask performed using the HEALPix package.

\section{The imprint of $E$-to-$B$ leakage and Minkowski functionals analysis}
\label{Minkowski Functionals as the statistic}

\subsection{Minkowski Functionals}
\label{Minkowski Functionals}

The MFs describe the morphological properties of convex, compact
sets in an $n$-dimensional space. They provide a powerful
statistical tool that can also be used in partial sky maps,
detecting non-gaussianities without previous knowledge of their
intensity or angular dependence
\cite{minkowski1903,gott1990,schneider1993,mecke1994,schmalzing1997,schmalzing1998,wizitzki1998}.
On a two-dimensional CMB field defined on the sphere,
$\mathcal{S}^2$, the morphological properties of the data can be
characterized as a linear combination of three MFs: the area,
contour length and integrated geodetic curvature of an excursion
set (the latter also known as the difference between the numbers
of hot and cold spots)\cite{gott1990,schmalzing1998}. For a
pixelized CMB sphere, an excursion set is given by the number of
pixels in which the temperature exceeds the threshold $\nu$. For a
given threshold $\nu$, it is convenient to define the excursion
set $Q_{\nu}$ and its boundary $\partial Q_\nu$ of a smooth scalar
field $u$ as follows,
\begin{eqnarray}
Q_{\nu}=\{x\in \mathcal{S}^2|u(x)>\nu\sigma\}, \\
\partial Q_{\nu}=\{x\in \mathcal{S}^2|u(x)=\nu\sigma\}.
 \end{eqnarray}

Thus, the area $v_0(\nu)$, the contour length $v_1(\nu)$ and the integrated geodetic curvature $v_2(\nu)$, can be written as \cite{schmalzing1998}:
 \begin{equation}
  \label{MFs}
 v_0(\nu)=\int_{Q_{\nu}}\frac{da}{4\pi},~v_1=\int_{\partial Q_{\nu}}\frac{d l}{16\pi},~v_2=\int_{\partial Q_{\nu}}\frac{\kappa dl}{8\pi^2},
 \end{equation}
where $da$ and $dl$ are the surface element of $\mathcal{S}^2$ and
the line element along $\partial Q_{\nu}$, respectively. The
geodetic curvature is represented by $k$. The MFs can be
numerically calculated for a given pixelized map $u(x_i)$ as
follows \cite{schmalzing1998,lim2012}:

 \begin{eqnarray}
 \label{MF0}
 v_0(\nu)=\frac{1}{N_{\rm pix}} \sum_{k=1}^{N_{\rm pix}} \Theta(u-\nu), \\
 \label{MF1_2}
 v_i(\nu)=\frac{1}{N_{\rm pix}} \sum_{k=1}^{N_{\rm pix}} \mathcal{I}_i(\nu,x_k),~~(i=1,2).
 \end{eqnarray}
In Eq. (\ref{MF0}), the Heaviside step function is represented by $\Theta$. In Eq. (\ref{MF1_2}), we define $\mathcal{I}$ such that

 \begin{eqnarray}
 \mathcal{I}_1(\nu,x_k)&=&\frac{\delta(u-\nu)}{4}\sqrt{u_{;\theta}^2+u_{;\phi}^2}, ~~~
 \mathcal{I}_2(\nu,x_k) = \frac{\delta(u-\nu)}{2\pi}\frac{2u_{;\theta}u_{;\phi}u_{;\theta\phi}-u_{;\theta}^2u_{;\phi\phi}-u_{;\phi}^2u_{;\theta\theta}}{u_{;\theta}^2+u_{;\phi}^2}.
 \end{eqnarray}
Expressing the covariant derivatives at a point $x= (\theta,
\phi)$, parameterized through the azimuth angle $\theta$ and the
polar angle $\phi$ of the unite sphere, we have
\cite{schmalzing1998}:
 \begin{gather}
u_{;\theta} = u_{,\theta}, ~~~
u_{;\phi} = \frac{1}{\sin \theta}  u_{,\phi}, ~~~
u_{;\theta \theta} = u_{,\theta \theta}, \\
u_{;\theta \phi} = \frac{1}{\sin \theta}u_{,\theta \phi} -\frac{\cos \theta}{\sin^2 \theta}u_{,\phi}, ~~~
u_{;\phi \phi} = \frac{1}{\sin^2 \theta}u_{,\phi \phi} + \frac{\cos \theta}{\sin \theta}u_{,\theta}.
  \end{gather}
In this paper, we shall investigate the statistical properties of
the $E$-to-$B$ leakage and its possible imprint  in the CMB
$B$-mode polarization field due to a partial sky analysis. We used
the algorithm developed by \cite{ducout2012} and \cite{gay 2012}
for calculating the MFs.

\subsection{Method}
\label{method}

%By comparing three different methods developed to extract the $E$ and $B$ signals from $Q$ and $U$, \cite{ferte2013} concluded that the Smith and Zaldarriaga method (SZ, hereafter) \cite{ebmixture3} is the most efficient since it significantly reduces the $E$-to-$B$ leakage, and at the same time ensures the smallest error bars. Following this result we will use the SZ method throughout this paper.

\begin{figure}[t]
\includegraphics[scale=0.3]{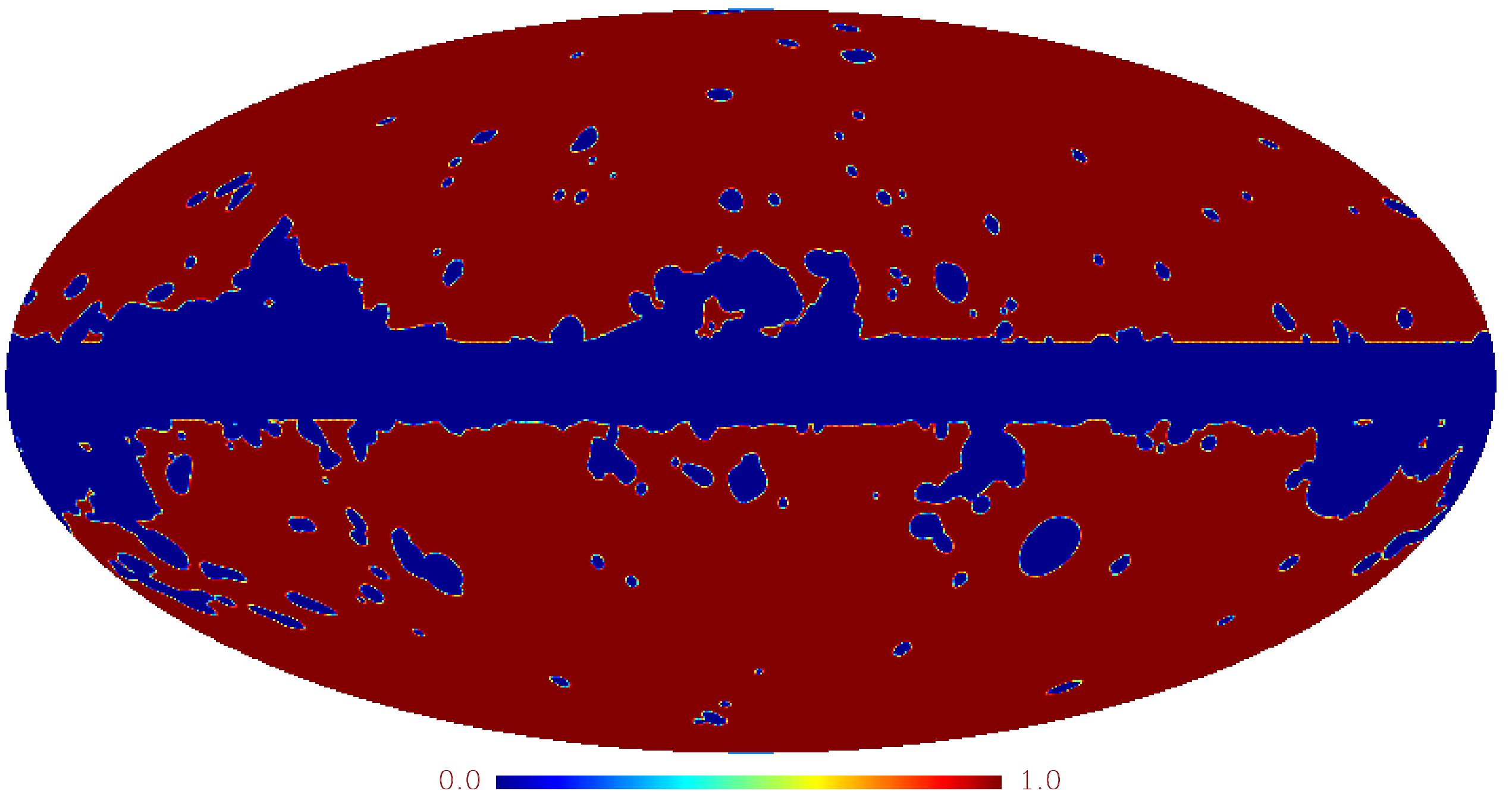}
\includegraphics[scale=0.3]{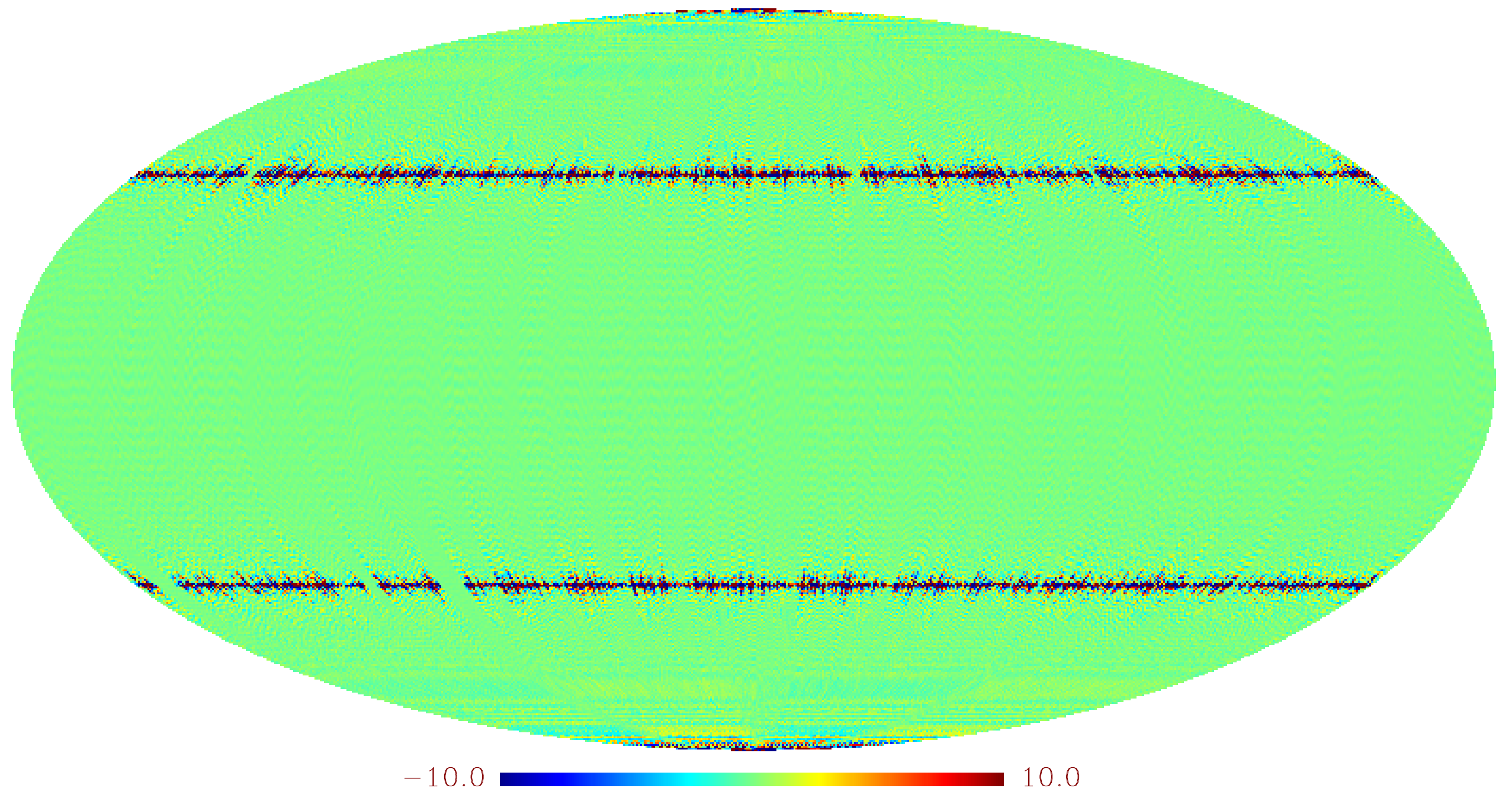}
\includegraphics[scale=0.3]{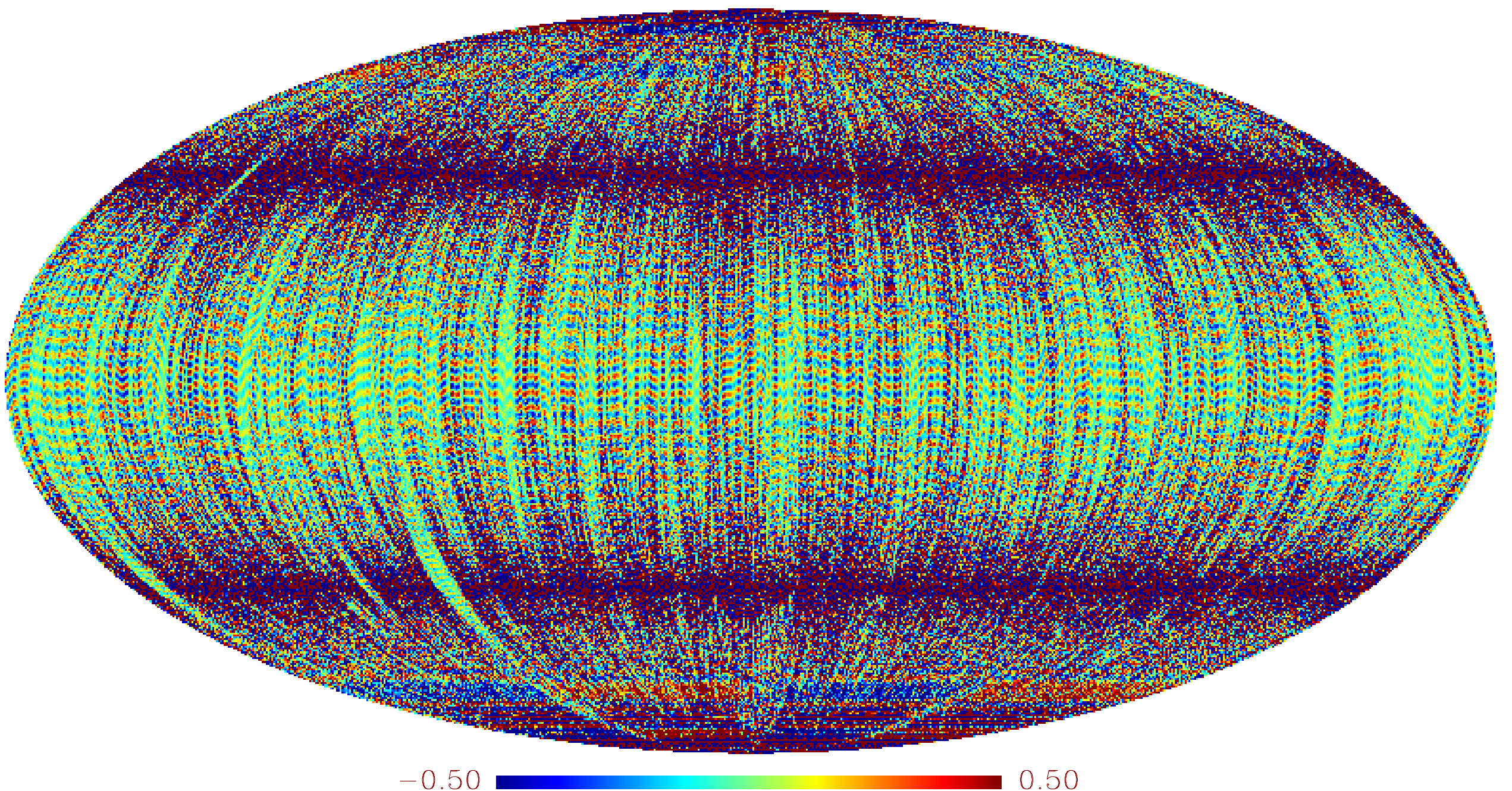}
\caption{Upper panel: On the left, the Gaussian smoothed window function considering Planck UT78 polarization mask with parameters $\beta= 10^{-4}$ and $\delta_c=1^{\circ}$. On the right, the residuals of $E$-to-$B$ leakage in $\mu$K for $r=0$ when CMB lensing is not taken into account. Lower panel: the same map shown in the right side of the upper panel rescaled.}
\label{mask_leakage}
\end{figure}

\begin{figure}[t]
\centerline{\includegraphics[scale=0.6]{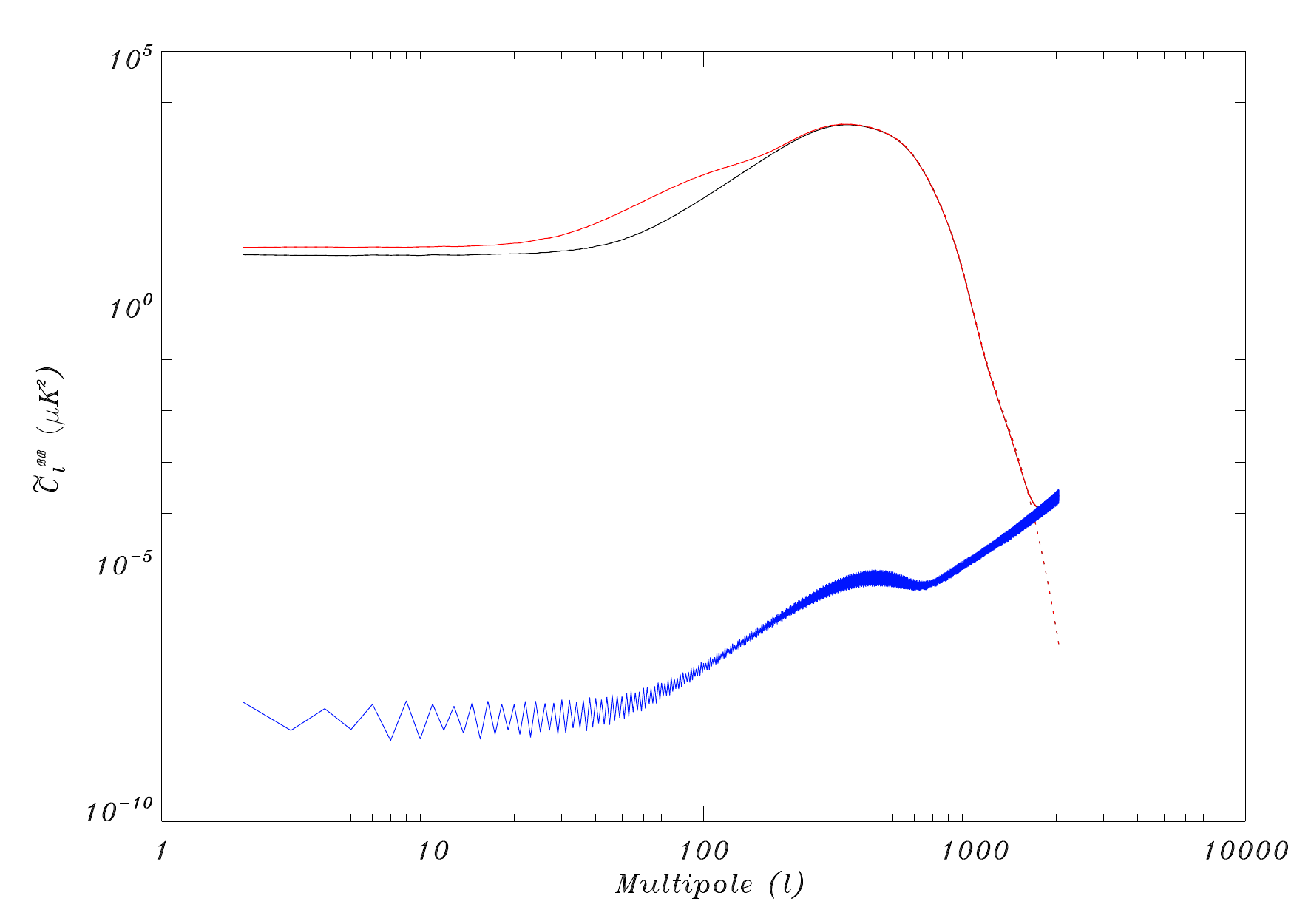}}
\caption{The mean power spectra of 500 simulations considering $r=0$ (black lines) and $r=0.1$ (red lines) for the ideal (dashed lines) and real (solid lines) cases. The blue solid curve is the mean power spectrum of 500 simulations considering $r=0$ and no CMB lensing for the real case (power spectrum related to the $E$-to-$B$ leakage). }
\label{ps}
\end{figure}

First, we generate two groups of Monte Carlo simulations
containing 500 full-sky $Q$ and $U$ lensed maps each. The first
group of simulation corresponds to a tensor-to-scalar ratio $r=0$,
and the second corresponds to $r=0.1$. We used the LensPix
software \cite{lenspix-paper} with cosmological parameters
$h^2\Omega_b=0.0223$,  $h^2\Omega_c=0.1188$, $h=0.673$,
$A_s=2.1\times10^{-9}$, $n_s=0.9667$, $\tau_{\rm reio}=0.066$ and
$N_{\rm side}=1024$. As mentioned above, in the numerical
calculations, the numerical errors caused by the high multipoles
are quite significant for two reasons: First, the errors caused by
the Gibbs phenomenon is dominant by the high multipoles
\cite{kim2011}; Second, the $\mathcal{B}\mathcal{B}$ power
spectrum is quite tilde blue, and the high multipoles dominate the
numerical errors in the HEALPix-based computations
\cite{ebmixture4}. Similar to the previous works
\cite{ebmixture4,Wang}, in order to smooth the high multipoles, we
apply a Gaussian smoothing with the parameter full width half
maximum ${\rm FWHM}=30'$.  We also smooth the edges of the Planck
polarization mask UT78pol, using Eqs. (\ref{WF}) and (\ref{betha})
with $\delta_c=1^{\circ}$ and $\beta =10^{-4}$  \cite{Wang} to
obtain our window function, shown in Fig. \ref{mask_leakage}. With
this in mind, we point out the steps of our analysis, which are
divided in two parts. In the first case (named as \emph{real case}
in this paper), in order to mimic the realistic data analysis, we
simulate the full-sky $Q$ and $U$ maps, and mask by adopting the
proper mask. Then, applying the $E$/$B$ decomposition method in
\cite{ebmixture3,ferte2013} we obtain the partial `pure' $B$-type
polarization map $\mathcal{B}(\hat{n})$. In the second case (named
as \emph{ideal case} in this paper), we derive the coefficients
$B_{\ell m}$ from the full-sky $Q$ and $U$ maps, and translate
them to the corresponding $\mathcal{B}_{\ell m}$. Thus, we can
construct the full-sky $\mathcal{B}$-maps by the standard route.
Then, we mask them by applying exactly same mask in the first
case. Comparing the $\mathcal{B}$-maps in these two cases, we find
that the maps derived in the {real case} include the residual
$E$-to-$B$ leakage caused by the $E$/$B$ decomposition method.
However, the maps derived from the {ideal case} are free from it
\cite{ebmixture4}. So, the difference between these two kinds of
$\mathcal{B}$-maps reflect the imprints of residual $E$-to-$B$
leakage, which is the main goal of this article. To realize it, we
do our analysis by the following steps:

\begin{itemize}

\item {\it Real case}: First, we obtain the $E$/$B$ decomposition
from partial $Q$ and $U$ lensed sky maps as numerically described
in \cite{ferte2013}. For each of the derived $\mathcal{B}W$-map,
we can define the pseudo power spectrum as
$\tilde{C}_{\ell}^{\mathcal{B}\mathcal{B}}=\frac{1}{2\ell+1}\sum_{m}
\tilde{\mathcal{B}}_{\ell m}\tilde{\mathcal{B}}^*_{\ell m}$, where
$\tilde{\mathcal{B}}_{\ell m}=\int \mathcal{B}W(\hat{n})Y^*_{\ell
m}(\hat{n}) d\hat{n}$. In Fig. \ref{ps}, we plot the average power
spectra $\tilde{C}_{\ell}^{\mathcal{B}\mathcal{B}}$ for the model
with $r=0$ and $r=0.1$ with solid black and blue lines,
respectively. As mentioned above, these power spectra include two
parts: One is the CMB $B$-type polarization, the other is the
residual $E$-to-$B$ leakage. In order to show the contribution of
the residual leakage, we do the exact same analysis to the model
with $r=0$ and no CMB lensing (i.e. no CMB $B$-mode). Thus, the
derived $\mathcal{B}W$-maps only include the residual $E$-to-$B$
leakage. From Fig. \ref{mask_leakage}, we find that the residuals
are quite significant around the two poles and two belts at
$\theta=48^{\circ}$ and $\theta=132^{\circ}$, which are caused by
the structure of the HEALPix package. The corresponding power
spectrum is also present in Fig. \ref{ps} (blue line), which is
much smaller than the spectra including CMB signals, in particular
in the low-multipole range. However, in the high-multipole range,
the residuals become more and more important, and dominate the
power spectrum at $\ell \gtrsim 1200$, which is consistent with
the previous works \cite{ebmixture3,ebmixture4,Wang}.

\item Second, it is well known that the calculation of the MFs
requires smoothing the maps to be analyzed in order to remove the
contribution of multipoles dominated by noise. Even though
different smoothing scales of the same CMB map have a high
correlation, they must be taken into account in order to extract
all its available statistical information. This is  based on the
fact that for each smoothing scale, the information of the CMB is
dominant in a different multipole range \cite{hikage2006,
hikage2008, ducout2012}. Thus, we smooth each final
$\mathcal{B}$-map with $N_{\rm side}=1024$ and ${\rm FWHM}=30'$
using a Gaussian filter with a smoothing scale, $\theta_s$, such
that $W_l=\exp \left[\frac{1}{2} \ell(\ell+1) \theta_s^2\right]$
with $\theta_s=10', 20', 30', 40', 50', 60'$, generating 6 sets of
500 maps.

\item Third, we apply three different sky cuts to the final
$\mathcal{B}$-maps: the smoothed (for each $\theta_s$) apodized
window function derived from the Planck UT78 polarization mask
(hereafter, sky cut 1), to exclude the pixels already without any
CMB information; the Planck mask + two bands centered at
$48^{\circ}$ and $132^{\circ}$, both with width of $6^{\circ}$ to
avoid the $E$-to-$B$ stripe residuals (hereafter, sky cut 2), and
Planck mask + the two residual bands + a $20^{\circ}$ width cut
around the poles, also to avoid the $E$-to-$B$ residuals in these
regions (hereafter, sky cut 3), see the upper panel right side of
Fig. \ref{mask_leakage}. For each smoothing scale, we excluded
every pixel of the window function with values less than 0.9 in
order to remove the boundary effects. We statistically analyze the
final 6 sets of 500 $\mathcal{B}$-map simulations for each sky cut
by means of the MFs for both groups with $r=0$ and $r=0.1$. Note
that the binning range of the threshold $\nu$ is set from $-3$ to
$3$ with 25 equally spaced bins.

 \item {\it Ideal case}: First, we obtain the $E$/$B$ decomposition
 in full-sky $Q$ and $U$ lensed sky maps directly using the HEALPix subroutine anafast. The $B_{\ell m}$ coefficients must be then multiplied by ${N_{\ell,2}}$ (see Eqs. (\ref{Elm_2}) and (\ref{Elm_pseudo})) before generating the full-sky $\mathcal{B}$-maps using the synfast subroutine of HEALPix. Then, we mask these $\mathcal{B}$-maps by applying the smoothed Planck UT78pol mask. Similar to the {real case}, we also calculate the corresponding power spectra $\tilde{C}_{\ell}^{\mathcal{B}\mathcal{B}}$, which are presented in Fig. \ref{ps} in dashed lines. Comparing with the results in the {real case}, we find that the spectra are same in both cases, which validates the effectiveness of the $E$/$B$ decomposition method.

\item  The second and third steps are exactly the same as for the real case.

 \item We finally
 compare the MFs of both cases, the one that the $E$-to-$B$ leakage is present (real) and the one that it is not (ideal),
 in order to identify the possible signature of the leakage in the MFs. We quantify the difference between the ideal and real
 cases by means of the $\chi^2$ statistics, defined as

 \begin{equation}
\label{snr}
\chi^2 = \sum_{aa'} \left[\bar{v}_a^{ideal} - \langle v_a^{real} \rangle \right] C^{-1}_{aa'} \left[\bar{v}_{a'} ^{ideal} - \langle v_{a'}^{real} \rangle \right],
\end{equation}

where $\langle \bar{v}_a^{real} \rangle $ is the model under test.
For each smoothing factor, $\theta_s$,  $a$ and $a'$ denote the
binning number of the threshold value $\nu$ and the different
kinds of MF. For the total $\chi^2_{T}$,  $a$ and $a'$ also denote
$\theta_s$. The covariance matrix is estimated from the average
under 500 simulations $C_{aa'} \equiv  \frac{1}{499}
\sum_{k=1}^{500} \left[\left(v_a^{k,real} -  \bar{v}_a^{real}
\right)\left(v_{a'}^{k,real} -  \bar{v}_{a'}^{real}
\right)\right]$.
\end{itemize}

\subsection{Results}
\label{Results}

In real CMB observations, the Galactic emission must be masked out
even considering data obtained from satellite surveys. The CMB
polarization, especially the $B$-mode signal, is the main target
of future experiments since it can probe inflation. However, this
primordial signal can be hidden behind the foregrounds:
instrumental noise, different astrophysical foregrounds and CMB
lensing (see \cite{larissa} for a detailed study of the statistics
of these different secondary $B$-mode signals). Moreover, dealing
with partial sky maps, leads to a leakage between $E$ and $B$
modes that could also mimic the primordial signal.

\begin{figure}[ht]
\centerline{\includegraphics[width=16cm,height=10cm]{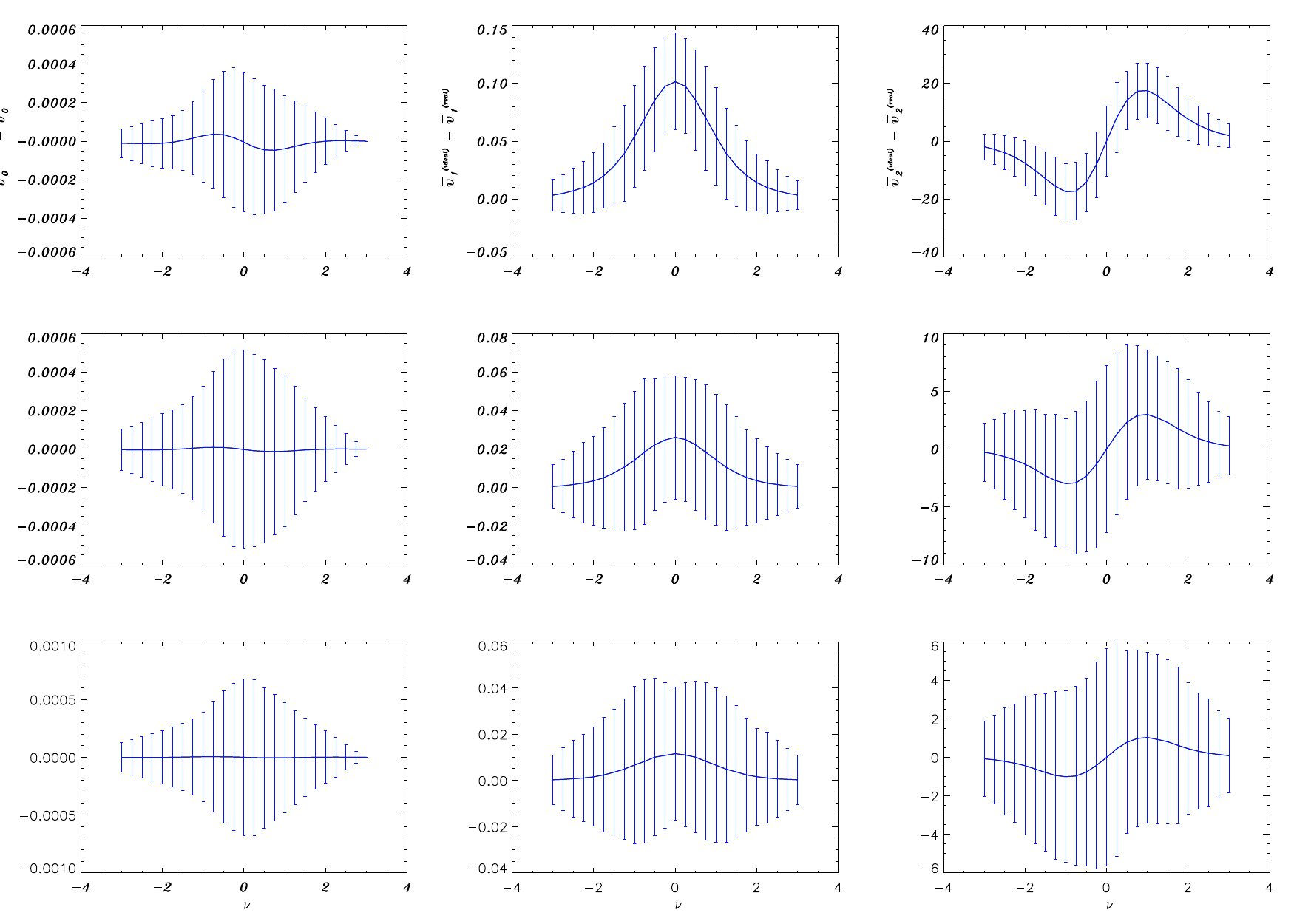}}
\caption{The difference between the mean values of real and ideal
case for the MFs considering $r=0$ over 500 simulations. From top
to bottom: $\theta_s=10', 40', 60'$, respectively. From left to
right: the first, second and third MF, respectively.}
\label{MF_r0}
\end{figure}

First, let's consider the case in which there is no primordial
gravitational waves, $r=0$. After proceeding with the $E$/$B$
decomposition, we applied to our simulations  the sky cuts
described in Sec. \ref{method} for each smoothing parameter,
$\theta_s$:  The UT78pol Planck mask alone, UT78pol +
contamination bands, and UT78pol + contamination bands + poles.
Our aim is to avoid both Galactic foreground and the $E$-to-$B$
leakage (check Fig. \ref{mask_leakage}). Calculating the MFs for
the final $\mathcal{B}$-map simulations and using Eq. (\ref{snr})
to obtain the $\chi^2$ statistics, we get the results shown in
Table \ref{sky_cut}. They quantify the significance of the leakage
for each smoothing scale and sky cut. We find that, by increasing
$\theta_s$ the significance of the leakage becomes much smaller,
which significantly shows that the leakage is dominated by the
high multipoles, which is consistent with the results of power
spectrum shown in Fig. \ref{ps}. We can see this result more clear
in Fig. \ref{MF_r0}, where we plotted $v_0, v_1, v_2$ (in Eq.
(\ref{MFs})) for $\theta_s=10',40', 60'$ in the case $r=0$. Even
though the significance of the leakage seen in Table \ref{sky_cut}
also decreases when we exclude the pixels where the leakage is
more evident, the contamination bands and the poles, its imprint
is still noticeable in the MFs analysis. We conclude that the
contamination bands and the poles do not play a very important
role in the overall leakage contribution, as we can also see in
the lower panel of Fig. \ref{mask_leakage}, where we rescaled the
unlensed $\mathcal{B}$-map with $r=0$ to point out the leakage
contribution.

\begin{table}[!htb]
\centering
\begin{tabular}{r r r r r r r}
\hline\hline
&  $\theta_s=10'$  & $\theta_s=20'$  &  $\theta_s=30'$  &  $\theta_s=40'$ &  $\theta_s=50'$  &  $\theta_s=60'$\\
 \hline
Sky cut 1 & 18.93 & 9.53  & 4.27 &1.62 & 0.71 & 0.34\\
Sky cut 2 & 17.27 & 8.55  & 3.80 &1.42 & 0.63 & 0.30 \\
Sky cut 3 & 15.78 & 7.97  & 3.56 &1.28 & 0.57 & 0.27 \\

\hline
\end{tabular}
\caption{$\chi^2$ for $r=0$ and three different sky cuts: the
UT78pol Planck mask alone, UT78pol + contamination bands, and
UT78pol + contamination bands + poles.} \label{sky_cut}
\end{table}

\begin{figure}[ht]
\centerline{\includegraphics[width=16cm,height=10cm]{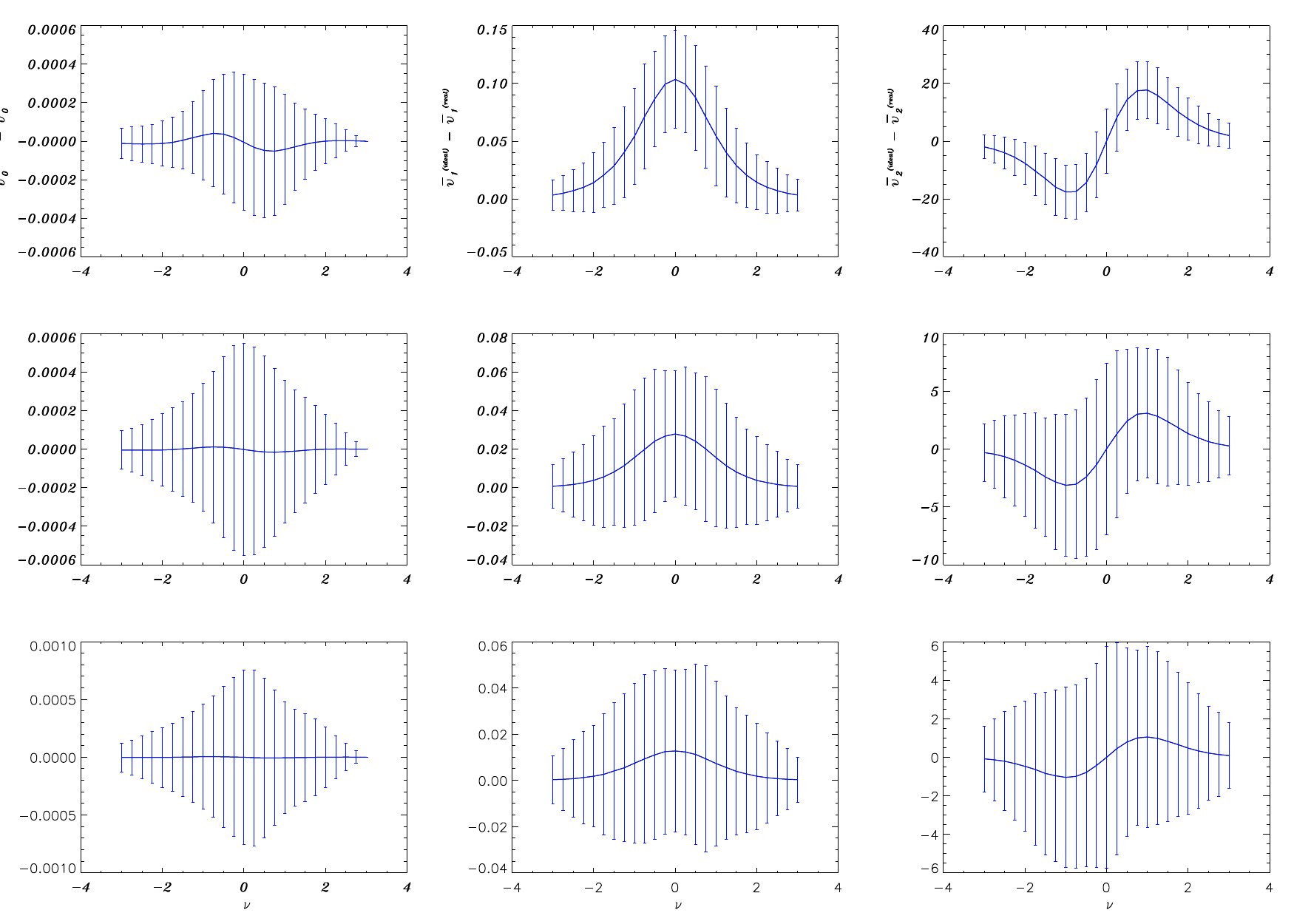}}
\caption{The difference between the mean values of real and ideal
case for the MFs considering $r=0.1$ over 500 simulations. From
top to bottom: $\theta_s=10', 40', 60'$, respectively. From left
to right: the first, second and third MF, respectively.}
\label{MF_r01}
\end{figure}

Now, considering both cases, without ($r=0$) and with primordial
gravitational waves ($r=0.1$), and applying the UT78 Planck
polarization mask only, we find no significant change in the
$\chi^2$ between them, as shown in Table \ref{r0_r1}. For the
larger tensor-to-scalar ratio, the effect of leakage is slightly
amplified for smaller $\theta_s$. We can compare the results for
the three MFs in Fig. \ref{MF_r0} with $\theta_s=10',40', 60'$
($r=0$) with the  same results for $r=0.1$ in Fig. \ref{MF_r01}.

\begin{table}[!htb]
\centering
\begin{tabular}{r r r r r r r}
\hline\hline
&  $\theta_s=10'$  & $\theta_s=20'$  &  $\theta_s=30'$  &  $\theta_s=40'$ &  $\theta_s=50'$  &  $\theta_s=60'$\\

\hline
$\chi^2$ $(r=0)$    & 18.93  &9.53    &4.27    &1.62  & 0.70 & 0.34 \\
$\chi^2$ $(r=0.1)$ & 21.18  &11.00  &4.41    &1.63  &0.62  &0.29\\
\hline
\end{tabular}
\caption{The $\chi^2$ for different smoothing parameters and
different models for the difference between ideal and real case,
considering only the UT78pol mask.} \label{r0_r1}
\end{table}

\begin{figure}[ht]
\includegraphics[scale=0.5]{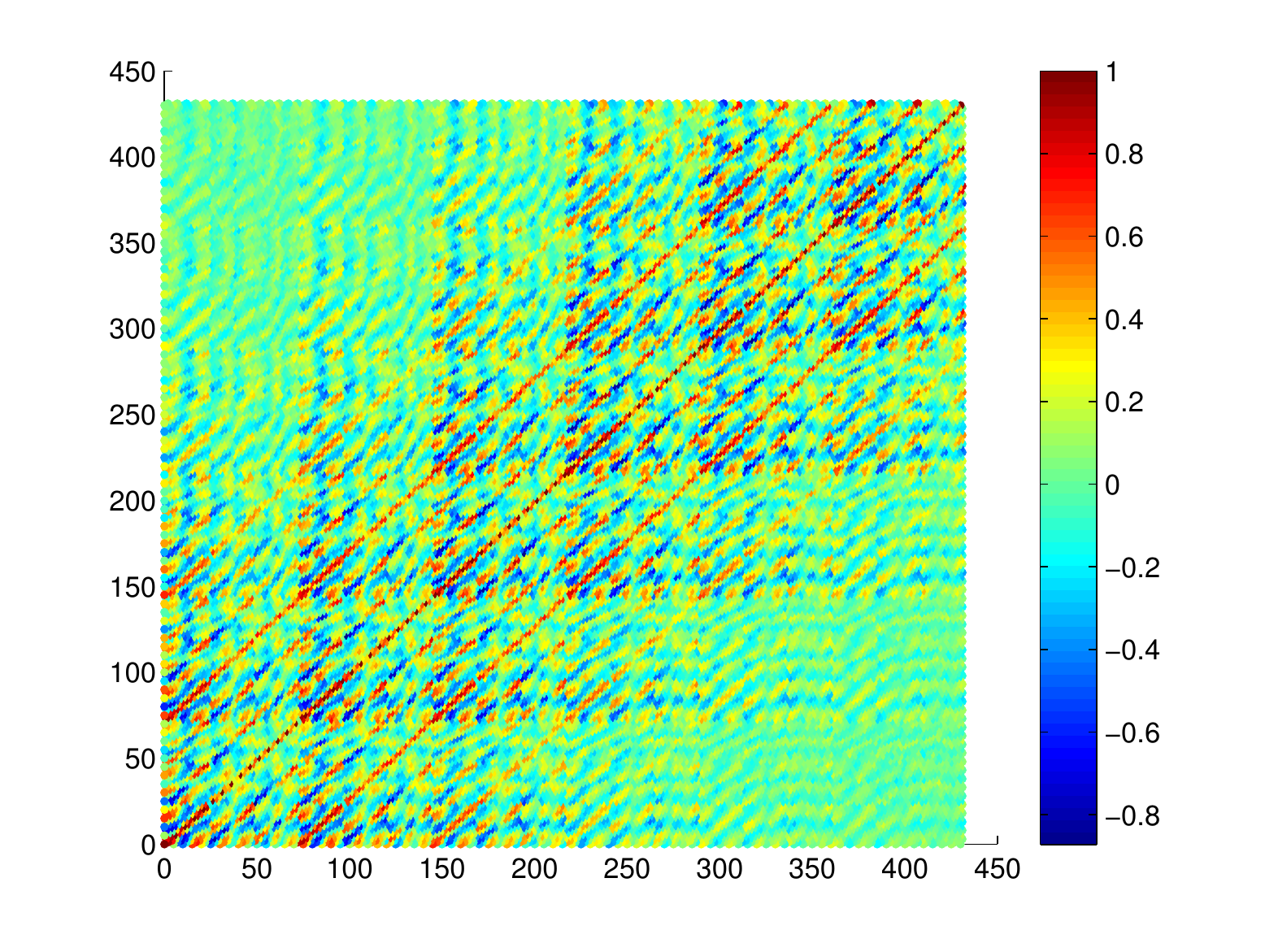}
\includegraphics[scale=0.5]{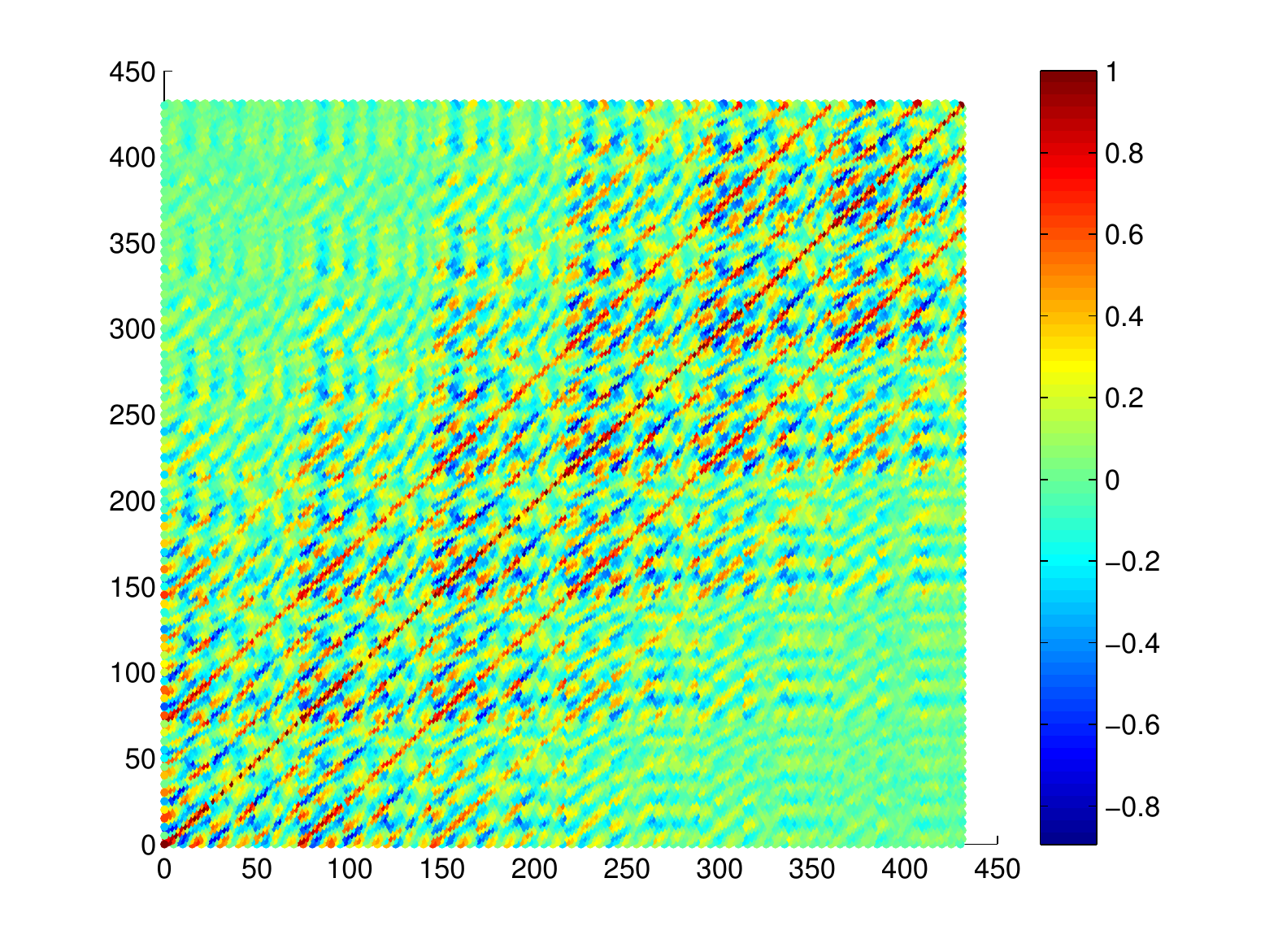}
\caption{The correlation coefficient values,
$\rho_{aa'}=C_{aa'}/\sqrt{C_{aa}C_{a'a'}}$,  for $r=0$ (left
panel), and for $r=0.1$ (right panel) are represented by the
colors. The axis correspond to $a$ and $a'$: the binning number of
the threshold value, the different kinds of MFs and the smoothing
scale.  Both panels correspond to the calculations when only the
UT78 Planck polarization mask is considered (sky cut 1).}
\label{cov}
\end{figure}

This latter result is more obvious when we analyze the total
$\chi^2$ combining all smoothing parameters in Eq. (\ref{snr}). We
can see in Table \ref{snr_t} that the effect of the $E$-to-$B$
leakage is more evident for $r=0.1$.  Comparing Tables \ref{r0_r1}
and \ref{snr_t}, we can also notice by the large values of the
total $\chi^2$ for both models that the MFs for different
smoothing scales are very correlated. It is easy to understand
this since the $E$-to-$B$ leakage is not a stochastic noise, and
it is always relevant in the same sky regions. The total $\chi^2$
is then obtained considering the correlation coefficients
$\rho_{aa'}=C_{aa'}/\sqrt{C_{aa}C_{a'a'}}$ instead of a direct sum
of the $\chi^2$ values for each smoothing scale, $\theta_s$. Note
that, in Fig. \ref{cov}, we did not use the last binning value of
the threshold $\nu$ since it approaches zero for the first MF for
every $\theta_s$ in order to avoid numerical problems. It is then
important to point out that even though the leakage seems not
relevant for individual smoothing scales, it is definitely
relevant when they are combined as shown in Table \ref{snr_t}. The
$E$-to-$B$ leakage should be carefully considered when analyzing
partial sky CMB $B$-mode data in order to avoid misinterpretation
of the data.

\begin{table}[!htb]
\centering
\begin{tabular}{r r r }
\hline\hline
 &  $r=0$  &  $r=0.1$\\
\hline
$\chi^2_T$ & 220.23  &268.74  \\
\hline
\end{tabular}
\caption{The total $\chi^2$ of MFs analysis for different models, considering only the UT78pol mask.}
\label{snr_t}
\end{table}

\section{Other statistics}
\label{Other statistics}

In order to further investigate the imprint of residual $E$-to-$B$
leakage in the constructed $\mathcal{B}$-maps, in this section, we
shall consider the other statistical tools (Betti numbers,
skewness, kurtosis), which are the complementary of analysis based
on MFs and power spectrum.

\subsection{Betti numbers}
\label{Betti numbers}

The morphological properties of the excursion sets can be also
quantified in terms of topological quantities called the Betti
numbers. They provide an intuitive understanding of the topology
of isosurfaces. For a two-dimensional manifold, such as the CMB
field, there are two non-zero Betti numbers. The excursion set
consists of many connected regions (number of hot spots)
$\beta_0$, and independent tunnels (number of cold spots)
$\beta_1$.  For each threshold $\nu$, we can mathematically
express $\beta_0$ and $\beta_1$ as line integrals
\cite{chingangbam 2012,Ganesan 2015}:
 \begin{eqnarray}
 \beta_0 =  \frac{1}{2\pi} \int_{C_+} k dl,  ~~~
 \beta_1 =  \frac{1}{2\pi} \int_{C_-} k dl,
 \end{eqnarray}
where $C_+$ and $C_-$ denote the contours that enclose hot and
cold spots, respectively. $k$ is the total curvature of
iso-temperature contours for each threshold $\nu$. Thus, the
genus, $g$, is given by a linear combination of $\beta_0$ and
$\beta_1$ as $g(\nu)=\beta_0(\nu)-\beta_1(\nu)$. The numerical
method for computing the Betti numbers is outlined in
\cite{chingangbam 2012,Ganesan 2015}.

Following the same steps outlined in Sec. \ref{method} and using
the exact same 500 $\mathcal{B}$-mode simulated maps, we
calculated $\beta_0$ and $\beta_1$ for both real and ideal cases
considering both $r=0$ and $r=0.1$. The imprint of the leakage in
the Betti number results were obtained by considering

 \begin{equation}
\Delta \beta_i = \beta_i^{ideal} - \beta_i^{real},
 \end{equation}
where $i=0,1$ represent the Betti numbers or alternatively, the number of hot and cold spots.

\begin{figure}[ht]
\includegraphics[scale=0.5]{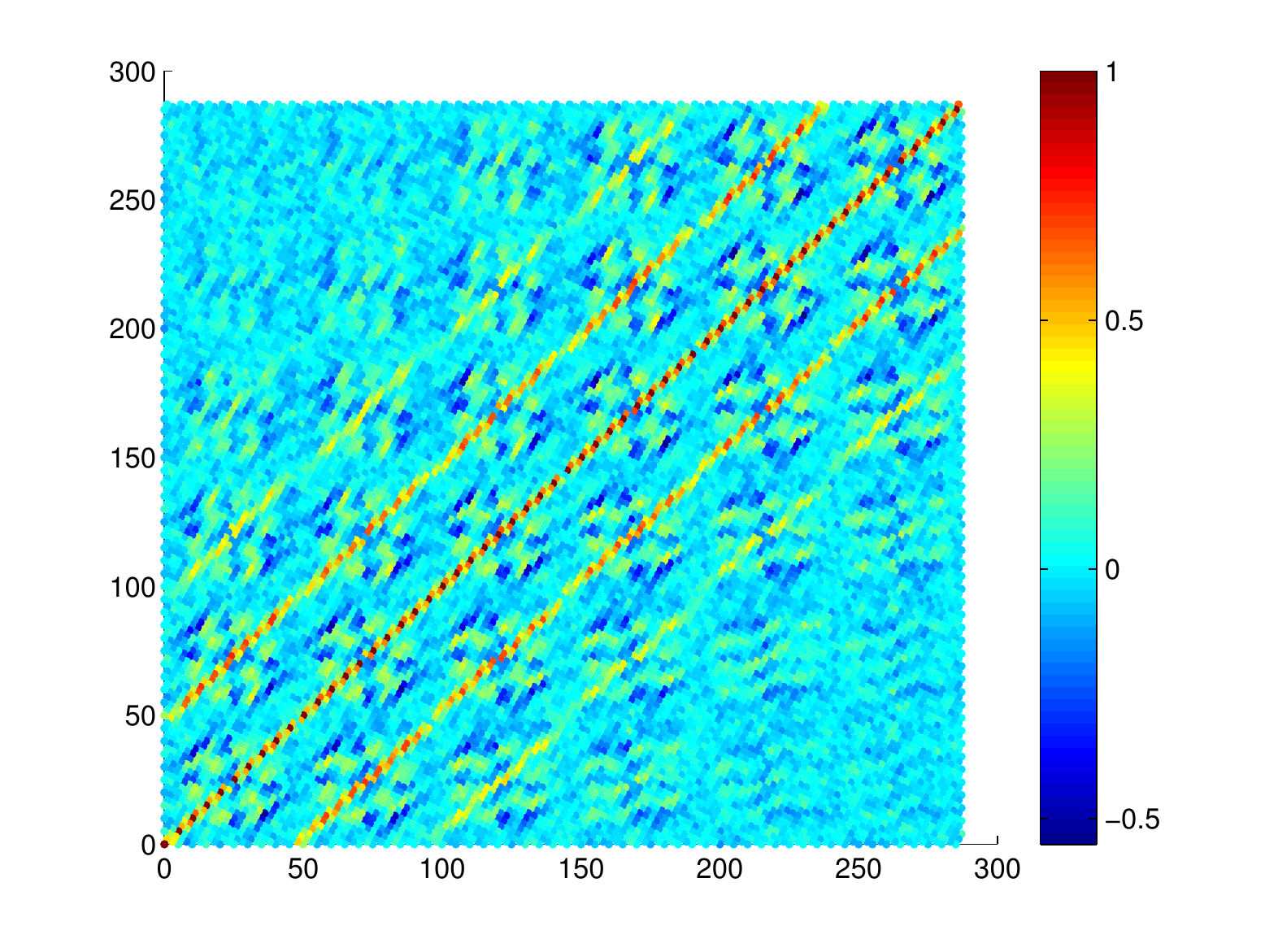}
\includegraphics[scale=0.5]{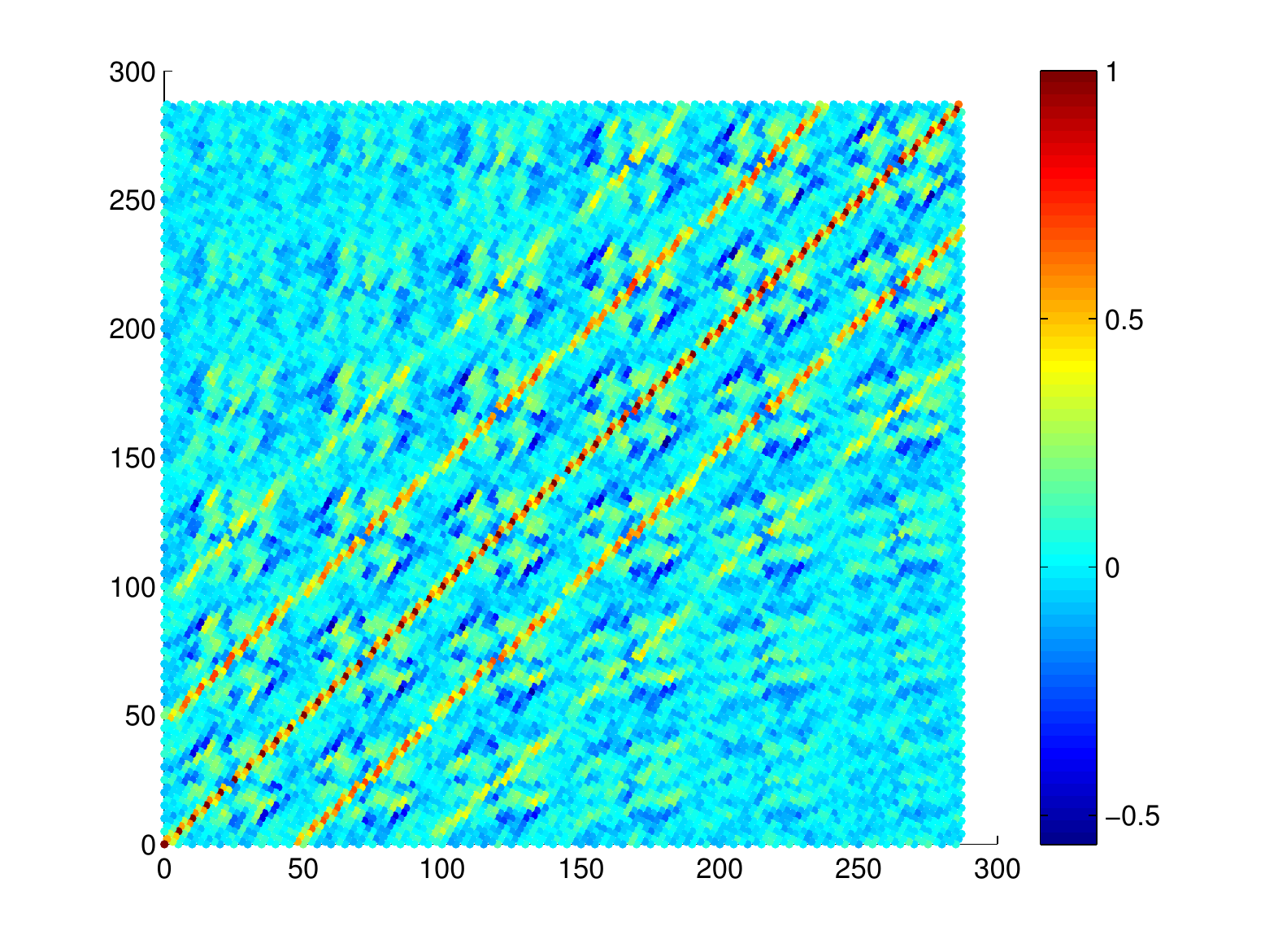}
\caption{The correlation coefficient values,
$\rho_{aa'}=C_{aa'}/\sqrt{C_{aa}C_{a'a'}}$,  for $r=0$ (left
panel), and for $r=0.1$ (right panel) are represented by the
colors. The axis correspond to $a$ and $a'$: the binning number of
the threshold value, the two Betti numbers  and the smoothing
scale.  Both panels correspond to the calculations when only the
UT78 Planck polarization mask is considered (sky cut 1).}
\label{cov_bn}
\end{figure}

\begin{figure}[ht]
\centerline{\includegraphics[width=16cm,height=10cm]{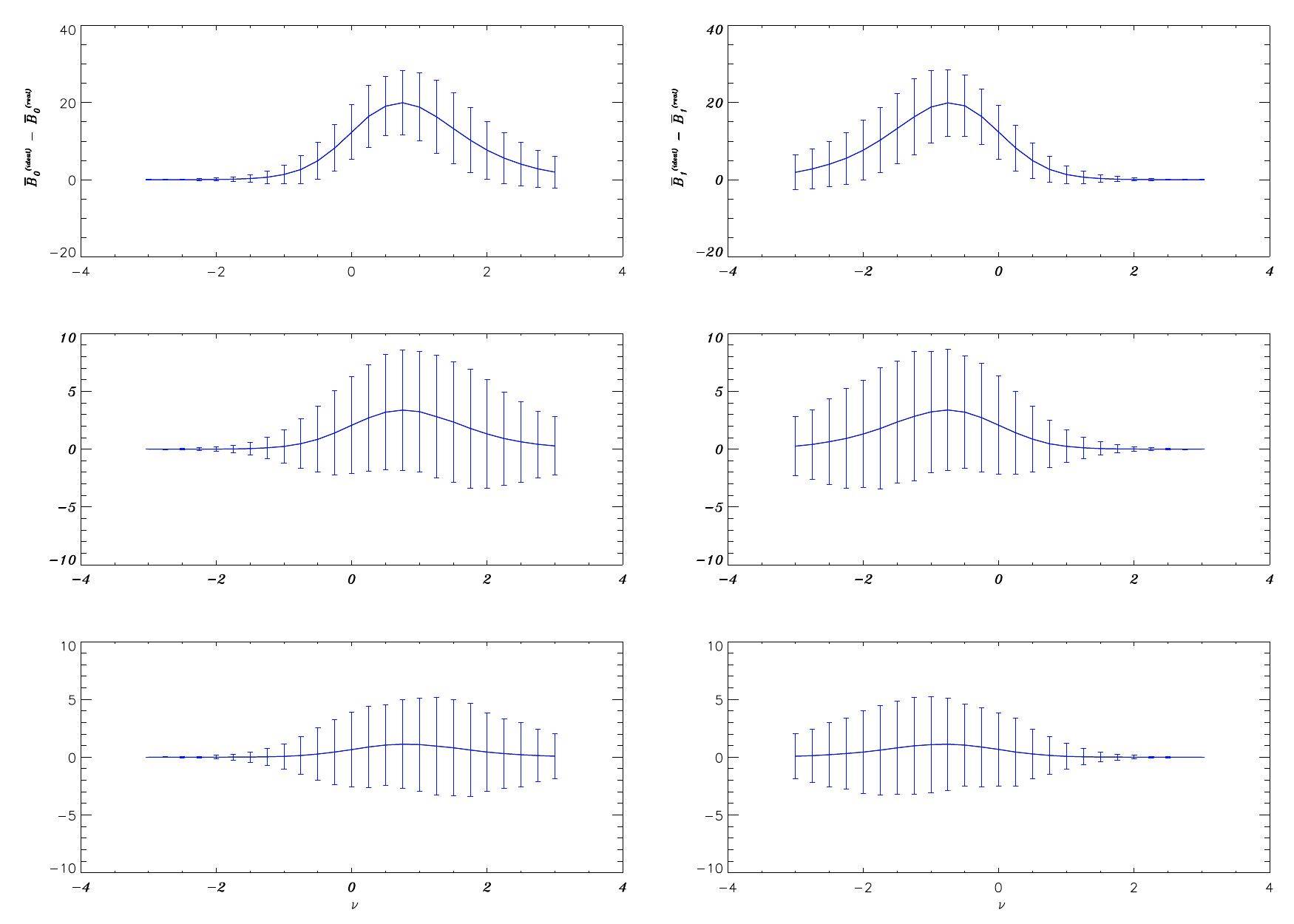}}
\caption{The difference between the mean values of real and ideal case for the Betti numbers considering $r=0$ over 500 simulations. From top to bottom: $\theta_s=10', 40', 60'$, respectively. From left to right: the first and the second Betti numbers, respectively.}
\label{BN_r0}
\end{figure}

\begin{figure}[ht]
\centerline{\includegraphics[width=16cm,height=10cm]{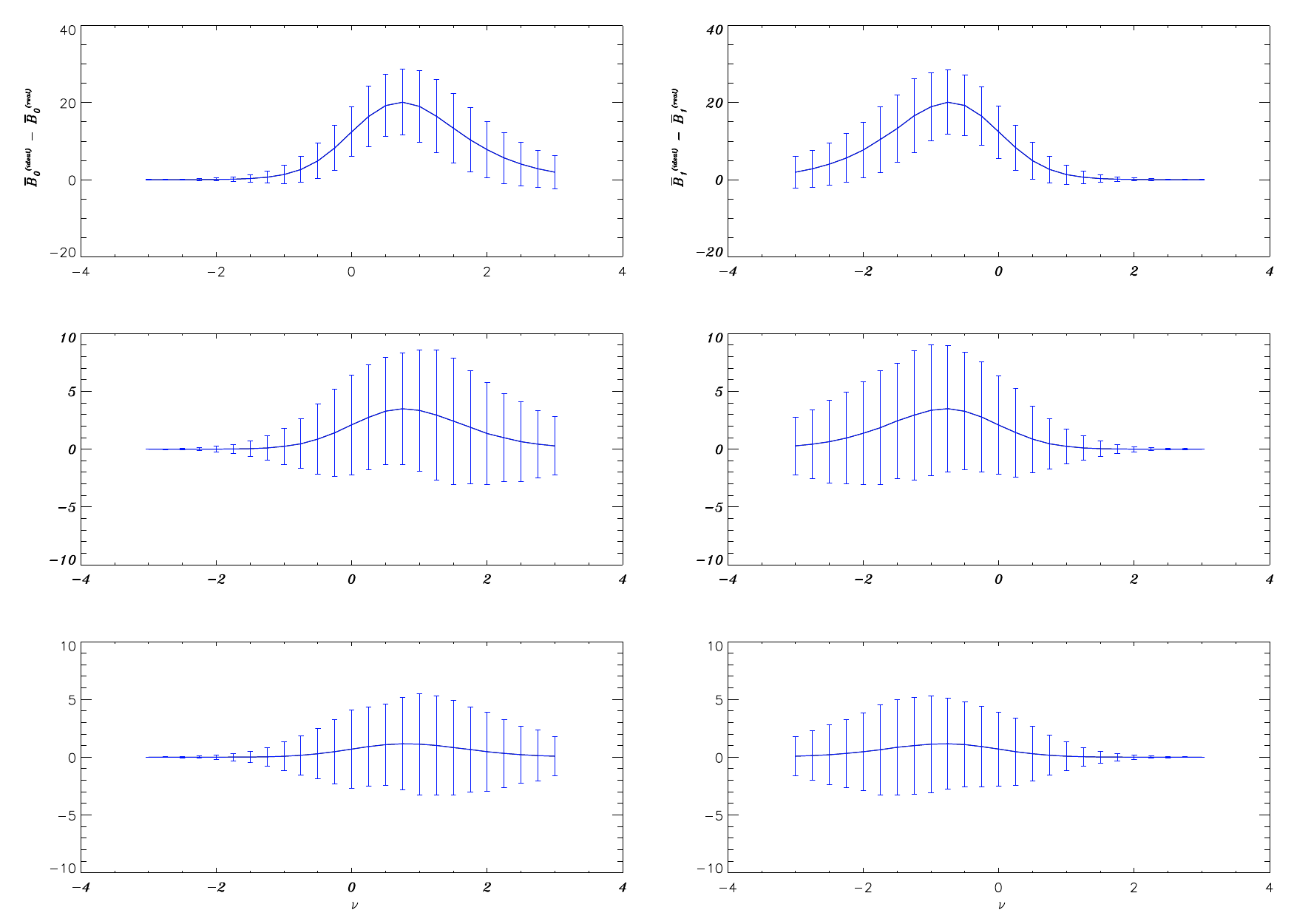}}
\caption{The difference between the mean values of real and ideal case for the Betti numbers considering $r=0.1$ over 500 simulations. From top to bottom: $\theta_s=10', 40', 60'$, respectively. From left to right: the first and the second, respectively.}
\label{BN_r01}
\end{figure}

To quantify  the difference between the ideal and real cases, we again use the $\chi^2$ statistics
 \begin{equation}
\label{snr}
\chi^2 = \sum_{aa'} \left[\bar{\beta}_a^{ideal} - \langle \beta_a^{real} \rangle \right] C^{-1}_{aa'} \left[\bar{\beta}_{a'} ^{ideal} - \langle \beta_{a'}^{real} \rangle \right].
\end{equation}
For each smoothing factor $\theta_s$,  $a$ and $a'$ denote the
binning number of the threshold value $\nu$ and the two Betti
numbers. We notice by Fig. \ref{BN_r0} (for $r=0$) and
\ref{BN_r01} (for $r=0.1$) that our results are qualitatively
consistent with the ones obtained for the MFs: The significance of
the leakage becomes smaller as $\theta_s$ increases. Moreover, we
found that the quantitative results of the $\chi^2$ statistics
considering each individual smoothing scale for the Betti numbers
are also in agreement with the results found for the MFs (compare
Tables \ref{sky_cut_bn} and \ref{r0_r1_bn} with \ref{sky_cut} and
Tables \ref{r0_r1}), supporting the method and providing a
consistency check for our calculations. Finally, as for the total
$\chi^2$, we see from Table \ref{snr_t_bn} that the imprint of the
$E$-to-$B$ leakage becomes less evident when compared with the
results for the MFs (Table \ref{snr_t}). It is important to
emphasize, however, that there is more information encoded in the
MFs analysis, especially considering all the correlation
coefficients when the smoothing scales are combined (compare Figs.
\ref{cov} and \ref{cov_bn}).  The calculation of the Betti numbers
is strongly related to the third MF $v_2$ \cite{chingangbam
2012,Ganesan 2015}.

\begin{table}[!htb]
\centering
\begin{tabular}{r r r r r r r }
\hline\hline
&  $\theta_s=10'$  & $\theta_s=20'$  &  $\theta_s=30'$  &  $\theta_s=40'$ &  $\theta_s=50'$  &  $\theta_s=60'$\\
 \hline
Sky cut 1 & 24.14 & 11.47   & 4.64 &1.84 & 0.74 & 0.35\\
Sky cut 2 & 22.31 & 10.45  & 4.17 &1.53 & 0.64 & 0.30 \\
Sky cut 3 & 19.17 & 9.11  & 3.93 &1.46 & 0.60 & 0.28 \\

\hline
\end{tabular}
\caption{$\chi^2$ for $r=0$ and three different sky cuts,
considering the Betti numbers: the UT78pol Planck mask alone,
UT78pol + contamination bands, and UT78pol + contamination bands +
poles.} \label{sky_cut_bn}
\end{table}

\begin{table}[!htb]
\centering
\begin{tabular}{r r r r r r r }
\hline\hline
&  $\theta_s=10'$  & $\theta_s=20'$  &  $\theta_s=30'$  &  $\theta_s=40'$ &  $\theta_s=50'$  &  $\theta_s=60'$\\
\hline
$\chi^2$ $(r=0)$    & 24.14  & 11.47  &4.64    &1.84 & 0.74 & 0.35 \\
$\chi^2$ $(r=0.1)$ & 23.70  &12.31  &5.12    &1.97  &0.75  &0.34\\
\hline
\end{tabular}
\caption{The $\chi^2$ for different smoothing parameters and
different models for the difference between ideal and real case
Betti numbers, considering only the UT78pol mask.}
\label{r0_r1_bn}
\end{table}

\begin{table}[!htb]
\centering
\begin{tabular}{r r r}
\hline\hline
 &  $r=0$  &  $r=0.1$\\
\hline
$\chi^2_T$ & 79.70  & 75.74  \\
\hline
\end{tabular}
\caption{The total $\chi^2$ of Betti numbers analysis for different models, considering only the UT78pol mask.}
\label{snr_t_bn}
\end{table}

\subsection{Skewness and kurtosis}
\label{Skewness and kurtosis}

In this subsection, for the cross-check, we apply the one-point
statistics, skewness and kurtosis to search for the imprints of
$E$-to-$B$ leakage. We calculated the skewness, the lack of
symmetry in a distribution, and the kurtosis, the degree to which
the distribution is peaked, for the 500 $\mathcal{B}$-mode
simulations considering both ideal and real cases and the 6
smoothing parameters mentioned previously ($\theta_s=10', 20',
30', 40', 50', 60'$).

\begin{figure}[t]
\centerline{\includegraphics[width=16cm,height=10cm]{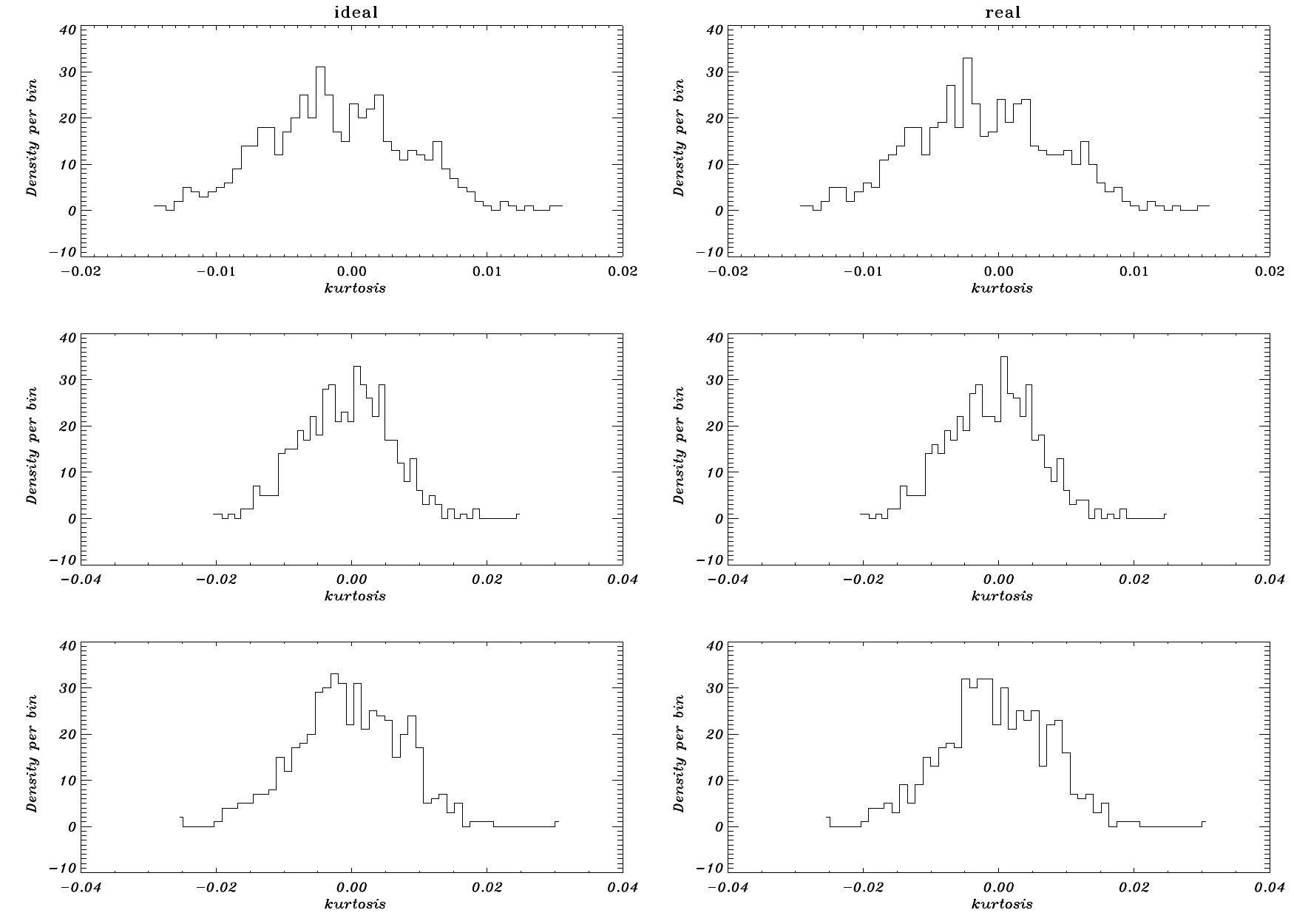}}
\caption{The value for the skewness for 500 simulations
considering $r=0$ for the ideal and real cases as specified in the
figure. From top to bottom: $\theta_s=10', 40', 60'$,
respectively. } \label{sqn_r0}
\end{figure}

\begin{figure}[t]
\centerline{\includegraphics[width=16cm,height=10cm]{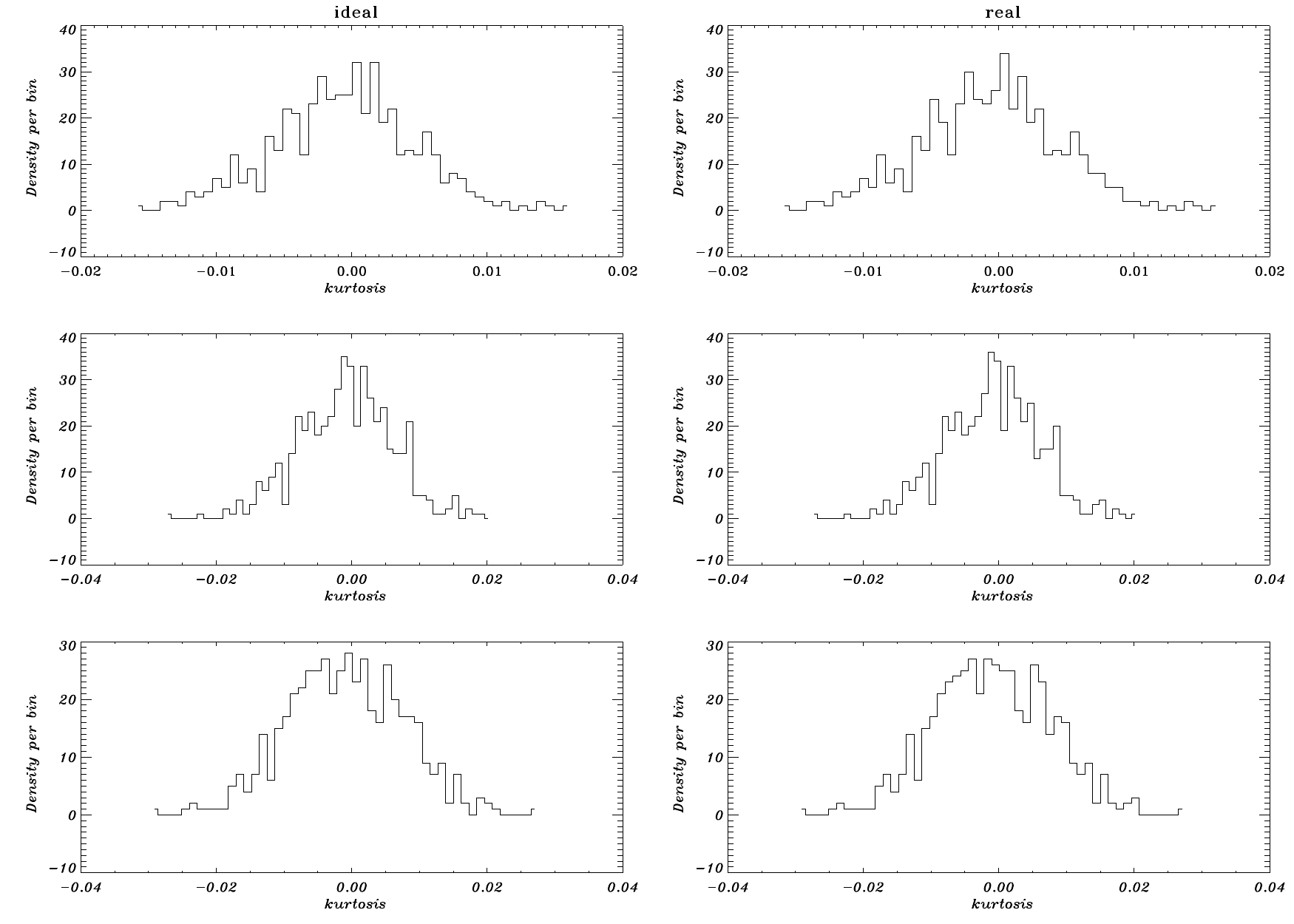}}
\caption{The value for the skewness for 500 simulations
considering $r=0.1$ for the ideal and real cases as specified in
the figure. From top to bottom: $\theta_s=10', 40', 60'$,
respectively. } \label{sqn_r01}
\end{figure}

\begin{figure}[t]
\centerline{\includegraphics[width=16cm,height=10cm]{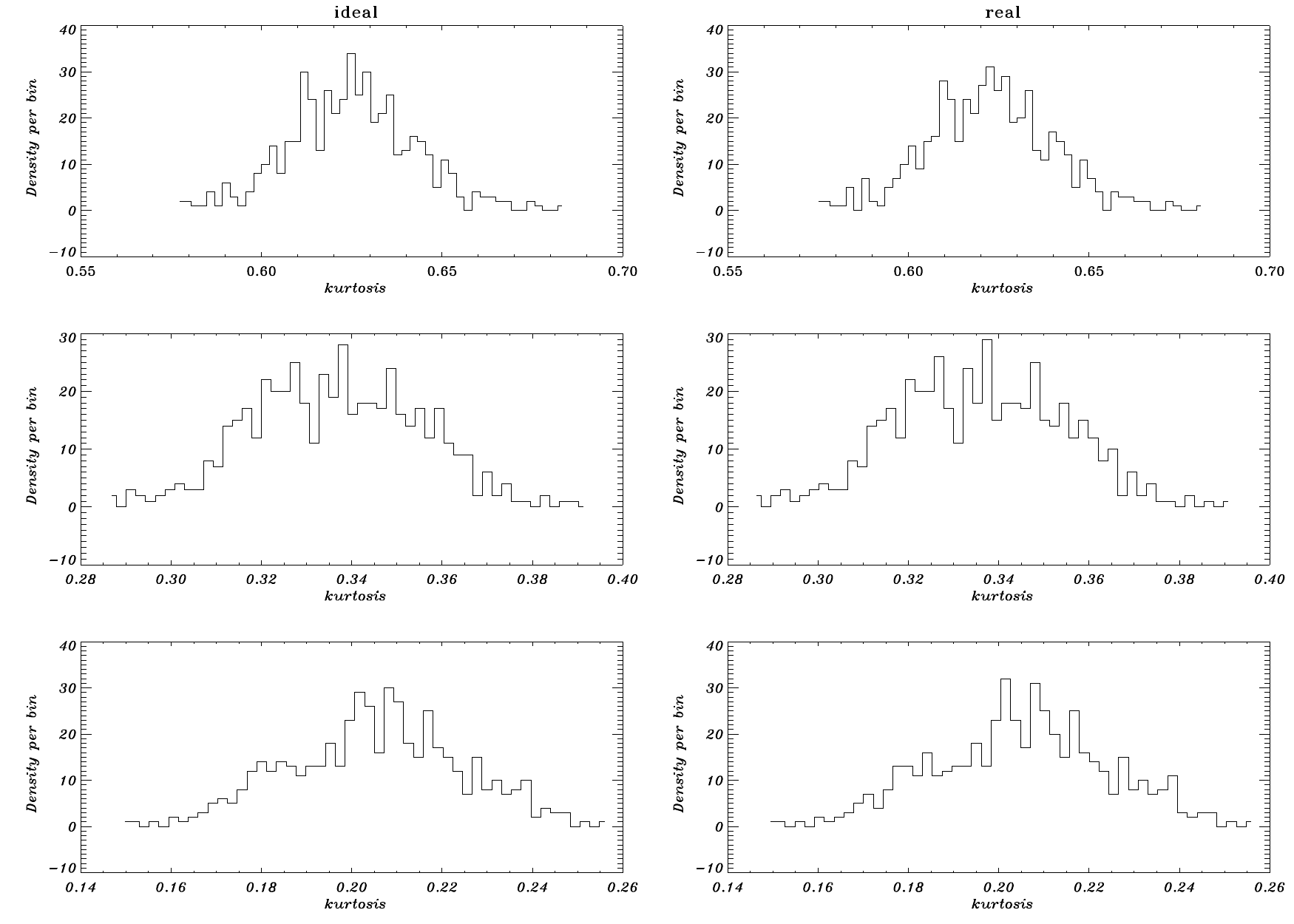}}
\caption{The value for the kurtosis for 500 simulations
considering $r=0$ for the ideal and real cases as specified in the
figure. From top to bottom: $\theta_s=10', 40', 60'$,
respectively. } \label{kurt_r0}
\end{figure}

\begin{figure}[t]
\centerline{\includegraphics[width=16cm,height=10cm]{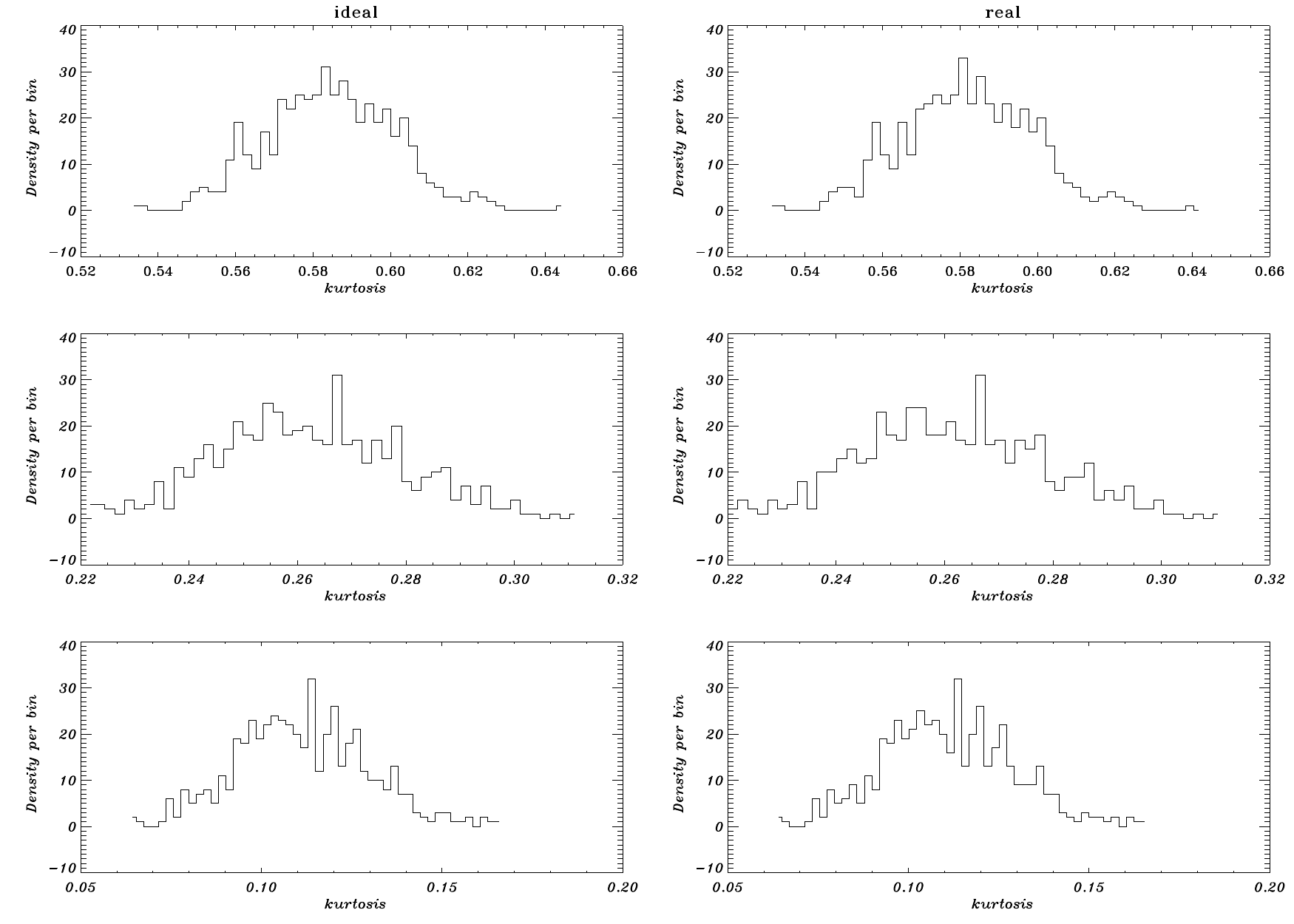}}
\caption{The value for the kurtosis for 500 simulations
considering $r=0.1$ for the ideal and real cases as specified in
the figure. From top to bottom: $\theta_s=10', 40', 60'$,
respectively. } \label{kurt_r01}
\end{figure}

The results in Tables \ref{skw} and \ref{kurt} show the mean
values and standard deviation for both statistics considering
$r=0$ and $r=0.1$, which show that no evidence of the leakage was
found neither in the skewness nor in the kurtosis statistics (we
see no deviations in the results comparing the ideal and real
cases). These analyzes indicate that, different from the MFs and
Betti numbers, the skewness and kurtosis statistics are not
sensitive enough to probe the imprints of residual $E$-to-$B$
leakage in the constructed $\mathcal{B}$-maps. The skewness
histograms of these simulations for the ideal and real cases
considering both $r=0$ and $r=0.1$ are shown in Figs. \ref{sqn_r0}
and \ref{sqn_r01}, respectively. The histograms were also plotted
for the kurtosis statistics, as can be seen in Figs. \ref{kurt_r0}
and \ref{kurt_r01}, for $r=0$ and $r=0.1$, respectively. These
figures are consistent with the results listed in Tables \ref{skw}
and \ref{kurt}, i.e. the differences between {ideal case} and real
case are not significant.

\begin{table}[!htb]
\centering
\resizebox{\columnwidth}{!}{%
\begin{tabular}{c @{\extracolsep{0.3em}} |r r r r r r}
\hline\hline
&  $\theta_s=10'~$  & $\theta_s=20'~$  &  $\theta_s=30'~$  &  $\theta_s=40'~$ &  $\theta_s=50'~$  &  $\theta_s=60'~$\\
\hline
$r=0$  & $~~-0.00057\pm0.00530$ &  $~~-0.00059\pm0.00560$ &  $~~-0.00040\pm0.00606$ & $~~0.00002\pm0.00669$  & $~~0.00040\pm0.00735$  &  $~~0.00058\pm0.00802$ \\
  & $~~-0.00057\pm0.00530$ &  $-0.00059\pm0.00561$ &  $-0.00039\pm0.00607$ & $0.00002\pm0.00670$  & $0.00040\pm0.00735$  &  $0.00058\pm0.00802$ \\
%\hline
$r=0.1$ & $0.00005\pm0.00525$ &  $0.00008\pm 0.00567$ &  $0.00013\pm0.00630$ & $0.00017\pm0.00705$  & $0.00018\pm0.00779$  &  $0.00021\pm0.00872$ \\
  & $~~0.00005\pm0.00526$ &  $~~0.00008\pm 0.00568$ &  $~~0.00013\pm0.00630$ & $~~0.00017\pm0.00705$  & $~~0.00018\pm0.00779$  &  $~~0.00021\pm0.00872$ \\
\hline
\end{tabular}
} \caption{The skewness values (mean value and the standard
deviation) for the simulated $\mathcal{B}$-mode polarization maps.
For each case, the upper one shows the results derived from the
ideal case, while the lower one shows those derived for the real
case. In both cases we use 500 lensed $\mathcal{B}$-maps
simulations and the UT78pol mask only.} \label{skw}
\end{table}

\begin{table}[!htb]
\centering
\begin{tabular}{c @{\extracolsep{0.3em}} |r r r r r r}
\hline\hline
&  $\theta_s=10'~$  & $\theta_s=20'~$  &  $\theta_s=30'~$  &  $\theta_s=40'~$ &  $\theta_s=50'~$  &  $\theta_s=60'~$\\
\hline
$r=0$  & $~~0.627\pm0.018$ &  $~~0.530\pm0.018$ &  $~~0.427\pm0.019$ & $~~0.338\pm0.019$  & $~~0.266\pm0.019$  &  $~~0.207\pm0.008$ \\
  & $0.625\pm0.018$ &  $~~0.529\pm0.018$ &  $~~0.426\pm0.019$ & $~~0.338\pm0.019$  & $~~0.266\pm0.019$  &  $~~0.207\pm0.008$ \\
%\hline
$r=0.1$ & $~~0.587\pm0.017$ &  $~~0.482\pm0.017$ &  $~~0.367\pm0.017$ & $~~0.264\pm0.017$  & $0.179\pm0.017$  &  $0.113\pm0.018$\\
        & $~~0.584\pm 0.017$ & $~~0.480\pm0.017$ &  $~~0.365\pm0.017$ & $~~0.263\pm0.017$  & $0.178\pm0.017$  &  $0.113\pm0.018$ \\
\hline
\end{tabular}
\caption{The kurtosis values (mean value and the standard
deviation) for the simulated $\mathcal{B}$-mode polarization maps.
For each case, the upper one shows the results derived from the
ideal case, while the lower one shows those derived for the real
case. In both cases we use 500 lensed $\mathcal{B}$-maps
simulations and the UT78pol mask only.} \label{kurt}
\end{table}

\section{Conclusions}
\label{Conclusions}

The CMB primordial $B$-type polarization is the main target of
future CMB observations since it provides the unique opportunity
to directly probe the evolution of the universe in the
inflationary stage. To avoid misunderstanding of the data, one
must take into account the different sources of the observed
$B$-mode signal, as for example, CMB weak lensing, astrophysical
foregrounds and instrumental noises. It is well known that the
unavoidable partial sky CMB analysis lead to a leakage from $E$ to
$B$ mode, acting as an extra noise for the primordial signal. In
this paper, we analyzed the effect of the $E$-to-$B$ leakage in
the measurement of the CMB $B$-mode. In order to clearly show the
effect of leakage residuals, and exclude the effects of the other
factors, throughout this paper, we do not consider the
instrumental noises (depending on the specific experiment, and the
survey scheme) and the possible foreground radiations.

A number of practical methods have been developed for the $E$/$B$
decomposition in the incomplete sky surveys. In this paper, we
adopted the method suggested by Smith et al., which ensures the
small error bars in all experimental configurations and leads to
the smallest leakage residuals. In addition, this method is based
on the algebraic framework of $\chi$-field, which avoids the high
computational cost and can be easily applied to the high
resolution CMB maps. However, even if this separation method is
used, the residual of the $E$-to-$B$ leakage is left in the
constructed $\mathcal{B}$-maps. In the present article, we
employed the MFs, Betti numbers, skewness and kurtosis statistics
to study the morphological imprint of the leakage in the
$\mathcal{B}$-mode polarization maps. We compared the ideal case,
in which the $\mathcal{B}$-maps were generated from full-sky $Q$
and $U$ observables, and the real case, where we generated the
$\mathcal{B}$-maps after applying the smoothed Planck polarization
mask UT78 to the $Q$ and $U$ sky by means of the Smith method
described in \cite{ebmixture3}. Different from the real case, the
ideal case is free from the $E$-to-$B$ leakage residual caused by
the $E$/$B$ separations. So, the difference between these two
cases reflects the imprint of $E$-to-$B$ leakage.

First, we compare the power spectra of $\mathcal{B}$-mode
polarization maps in these two cases, and found that they are
nearly same, in particular in the low-multipole range, and the
contribution of $E$-to-$B$ leakage is tiny. Second, we did not
find any imprint of the $E$-to-$B$ leakage in the simulated
$\mathcal{B}$-maps when considering both skewness and kurtosis
statistics, which shows that these one-point statistics are not
sensitive to the leakage. These results confirm the effectiveness
of the $E$/$B$ separation method.

However, by comparing the MFs and Betti numbers applied to the
$\mathcal{B}$-mode maps generated in the ideal and real cases, we
find that the effect of leakage residuals decreases quickly with
the increasing of $\theta_s$, which significantly shows that the
residuals are dominant by the higher multipoles. This result is
consistent with the one derived from the power spectrum analysis.
In addition, we also found that the leakage cannot be ignored when
combining the results of all the smoothing scales, $\theta_s$, in
both models, $r=0$ and $r=0.1$. Considering individual smoothing
scales leads to the mistaken conclusion that the significance of
the leakage is small and that it can be safely neglected. It is
then important to point out that the large correlation of the MFs
for different smoothing scales is expected since the leakage is
not randomly distributed in the sky. The $E$-to-$B$ leakage plays
an important role in the final $\mathcal{B}$-map and must be taken
into account to avoid misinterpretation of the data.

~

~

\section*{Acknowledgements}
We acknowledge the use of the Planck Legacy Archive (PLA). Our
data analysis made the use of HEALPix \cite{healpix}, CAMB
\cite{camb} and LensPix \cite{lenspix}. This work is supported by
NSFC No. J1310021, 11603020, 11633001, 11173021, 11322324,
11653002, project of Knowledge Innovation Program of Chinese
Academy of Science and the Fundamental Research Funds for the
Central Universities.

\end{document}